\documentclass[english,showkeys,nofootinbib,notitlepage]{revtex4-1}
\usepackage[T1]{fontenc}
\usepackage[latin9]{inputenc}
\usepackage{babel}
\usepackage{amsmath}
\usepackage{amssymb}
\usepackage{graphicx}
\usepackage{esint}
\usepackage[unicode=true,pdfusetitle,
 bookmarks=true,bookmarksnumbered=false,bookmarksopen=false,
 breaklinks=false,pdfborder={0 0 1},backref=false,colorlinks=false]
 {hyperref}

\makeatletter

\providecommand{\tabularnewline}{\\}

\usepackage{babel}

\usepackage{babel}
\usepackage{babel}

\usepackage{babel}

\makeatother

\begin{document}

\title{Production mechanisms of open-heavy flavor mesons}

\author{Iván Schmidt, Marat Siddikov}

\affiliation{Departamento de Física, Universidad Técnica Federico Santa María,~~\\
 y Centro Científico - Tecnológico de Valparaíso, Casilla 110-V, Valparaíso,
Chile}
\begin{abstract}
In this paper we discuss different mechanisms of open-heavy flavor
meson production. Using the color dipole framework, we analyze in
detail the contributions of the conventional two-pomeron fusion and
the three-pomeron fusion correction. In a parameter-free way we found
that the three-pomeron mechanism is significant for $D$-meson production in
the small-$p_{T}$ kinematics, although it is less important at large $p_{T}$,
as well as for $B$-mesons. The inclusion of the three-pomeron mechanism
significantly improves the agreement of theoretical predictions with experimental
data in the small-$p_{T}$ kinematics. We also consider 
the non-prompt charmonia production, and demonstrate that the theoretical
results are in reasonable agreement with experimental data. Finally,
we compare the theoretical predictions for the dependence on multiplicity
of co-produced hadrons to experimental data recently measured by the ALICE
collaboration. We found that, contrary to naive expectations, the contribution
of the three-pomeron mechanism has only a mild effect on the self-normalized
observables in the range of multiplicities studied at ALICE, and for this
reason the two-pomeron fusion mechanism can describe reasonably well
the experimentally observed multiplicity dependence. 

\end{abstract}
\maketitle

\section{Introduction}

Hadrons containing heavy quarks present a widely used tool to
test the predictions of Quantum Chromodynamics (QCD). In the heavy
quark mass limit the dynamics of a heavy quark can be described
perturbatively~\cite{Korner:1991kf}, which allows to test the perturbative
QCD (pQCD) predictions. For this reason the production of heavy
mesons has been extensively studied in the literature~(see e.g. ~\cite{Bodwin:1994jh,Maltoni:1997pt,Binnewies:1998vm,Kniehl:1999vf,Brambilla:2008zg,Feng:2015cba,Brambilla:2010cs,Ma:2018bax,Goncalves:2017chx}
for overview), and a reasonable description of the experimental data
on inclusive production has been achieved. However, the existing theoretical
models are constantly being challenged by the improving precision of the
available data and the technical advances which make it possible
to measure more complicated observables. In fact, the start of the High Luminosity
Run 3 at LHC (HL-LHC mode)~\cite{ATLAS:2013hta,Apollinari:HLLHC,LaRoccaRiggi}
will significantly enhance the available data and will give the possibility
to analyze the mechanisms of different processes.

One of the observables which can be measured, thanks to the large luminosity,
is the dependence of the cross-section on yields (multiplicity) of
the charged particles co-produced together with a given heavy meson~\cite{Adam:2015ota,Trzeciak:2015fgz,Ma:2016djk,PSIMULT,Khatun:2019slm,Alice:2012Mult}.
Since the charged particles are produced nonperturbatively, this observable
allows to test an interplay of the soft and hard physics, while, as was explained
in~\cite{Siddikov:2019xvf,Levin:2018qxa}, the high multiplicity
events allow to test the physics in a deeply saturated regime, which
otherwise would require significantly larger energies. Recent experimental
data~\cite{Trzeciak:2015fgz,Ma:2016djk,PSIMULT,Khatun:2019slm,Alice:2012Mult}
show that the yields of $S$-wave quarkonia ($J/\psi$, $\psi(2S),\,\Upsilon(1S)$)
grow rapidly as a function of the multiplicity of charged particles. Such
behavior is at tension with conventional two-pomeron fusion mechanisms,
and potentially might signal that there are sizable contributions
from three-pomeron fusion mechanisms~\cite{Siddikov:2019xvf,Levin:2018qxa},
which were previously disregarded as higher twist effects. For this
reason it is important to revisit the analysis of open heavy-flavor
($D$- and $B$-) meson production and check if the conventional mechanisms
can describe the multiplicity dependence. In the case of $D$- mesons
such dependence was recently measured by the ALICE collaboration~\cite{Adam:2015ota},
while in $B$-meson production to the best of our knowledge there is no
direct experimental data on the multiplicity dependence~\footnote{$B$-mesons are usually reconstructed from the $B\to J/\psi\,K$ decay
channel, which has an order of magnitude smaller branching fraction
than the inclusive $B\to J/\psi+X$ decay. For this reason it is more
challenging to study this channel, especially in rare high-multiplicity
events.}, yet there are data on the multiplicity dependence of non-prompt
$J/\psi$ production, which proceeds via $B\to J/\psi$ decays~\cite{Adam:2015ota}.
These data allow to test the predicted multiplicity dependence for
the case of heavier $b$-quarks. The range of multiplicites in the
currently available experimental data is quite limited, because the
statistics falls rapidly as a function of event multiplicity, but
we expect that it will be significantly extended with data from Run
3 at LHC (HL-LHC mode)~\cite{ATLAS:2013hta,Apollinari:HLLHC,LaRoccaRiggi}.
Another goal of this paper is to estimate the contribution of the
three-pomeron mechanisms, which are usually disregarded in the heavy
quark mass limit. While in the dipole picture it is believed that a
universal dipole cross-section should take into account all such contributions,
in phenomenological parametrizations of the
dipole cross-section usually such contributions are taken into
account only partially or not taken into account at all. For this
reason in our explicit evaluation we estimate explicitly the role
of such contributions. In particular, since the contribution of the three-pomeron
mechanism is expected to grow faster than that of the two-pomeron
fusion, we pay special attention to the three-pomeron mechanism in
large multiplicity events.

The paper is structured as follows. In Section~\ref{sec:Evaluation}
we discuss the framework used for the evaluation of the open-heavy meson
production. In Subsection~\ref{subsec:2Pom} we evaluate the contribution
of the two-pomeron fusion mechanism and compare its predictions with
experimental data. In Subsection~\ref{subsec:3Pom-1} we evaluate
the contributions of the three-pomeron mechanisms and estimate numerically
their relative contributions. Our major finding is that they are significantÄ
for $D$-mesons for small $p_{T}\lesssim10$ GeV, yet become negligible
for large $p_{T}$ and for $B$-mesons. In Section~\ref{sec:Numer}
we develop the framework for the multiplicity dependence description 
in the dipole formalism and compare its predictions for the multiplicity dependence
with available experimental data. Finally, in Section~\ref{sec:Conclusions}
we draw conclusions.

\section{Evaluation of the inclusive cross-section}

\label{sec:Evaluation}In this section we will focus on the production
of open heavy-flavor mesons ($D$- and $B$-mesons). The cross-section
for heavy meson production can be related to the cross-section
for heavy quark  production as~\cite{Binnewies:1998vm,Kniehl:1999vf,Ma:2018bax,Goncalves:2017chx}.
\begin{equation}
\frac{d\sigma_{pp\to M+X}}{dy\,d^{2}p_{T}}=\sum_{i}\int_{x_{Q}}^{1}\frac{dz}{z^{2}}D_{i}\left(\frac{x_{Q}(y)}{z}\right)\,\frac{d\sigma_{pp\to\bar{Q}_{i}Q_{i}+X}}{dy^{*}d^{2}p_{T}^{*}}\label{eq:fragConvolution}
\end{equation}
where $y$ is the rapidity of the heavy meson ($D$- or $B$-meson),
$y^{*}=y-\ln z$ is the rapidity of the heavy quark, $p_{T}$ is the
transverse momentum of the produced $D$-meson, $D_{i}(z)$ is the
fragmentation function which describes fragmentation of the parton
$i$ into a heavy meson, and $d\sigma_{pp\to\bar{Q}_{i}Q_{i}+X}/dy^{*}$
is the cross-section of heavy quark production with rapidity $y^{*}$.
The fragmentation functions for the $D$- and $B$-mesons, as well
as non-prompt $J/\psi$ production, are known from the literature (see
the Appendix~\ref{sec:FragFunctions} for details). Since the dominant
contribution to all mentioned states stems from the heavy $c$- and
$b$-quarks (prompt and non-prompt mechanisms respectively),  their
production can be evaluated in the heavy quark mass limit, and for this
reason in what follows we will focus on the evaluation of the cross-section
$d\sigma_{pp\to\bar{Q}_{i}Q_{i}+X}/dy^{*}d^{2}p_{T}^{*}$, which appears
in the integrand of~(\ref{eq:fragConvolution}).

\subsection{Two-pomeron contribution}

\label{subsec:2Pom}
\begin{figure}
\includegraphics[width=9cm]{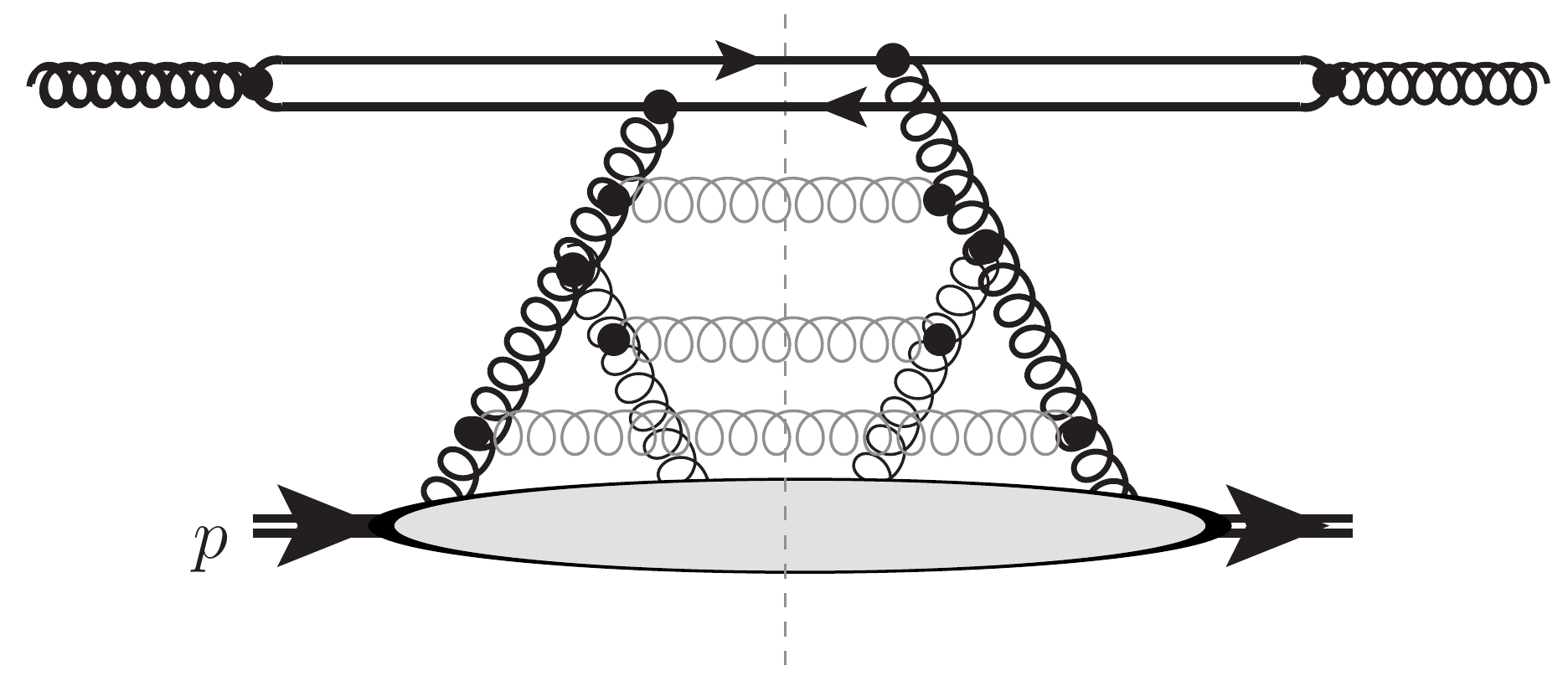}

\caption{\label{fig:DipoleCrossSections-2Pom}A typical two-pomeron fusion
diagram, taken into account in evaluation of the heavy quark production.
In the dipole framework~\cite{Iancu:2003ge,RESH,Kowalski:2006hc,Watt:2007nr}
the dipole cross-section is found as a solution of the Balitsky-Kovchegov
(BK) equation, so effectively the BK pomeron includes additional fan-like
contributions (shown by grey lines; resummation of all possible fan-like
topologies is implied). The vertical dashed grey line stands for
the unitarity cut, the blob in the lower part is the hadronic target
(proton); the fermionic loop in the upper part of the figure includes
a summation over all possible gluons. }
\end{figure}
The conventional mechanism widely used for description of the heavy
meson production is a pomeron-pomeron fusion (see Figure~\ref{fig:DipoleCrossSections-2Pom}).
The corresponding cross-section in the dipole approach is given by~\cite{Ma:2018bax,Goncalves:2017chx}
\begin{eqnarray}
 &  & \frac{d\sigma_{pp\to\bar{Q}_{i}Q_{i}+X}\left(y,\,\sqrt{s}\right)}{dy\,d^{2}p_{T}}=\,\int d^{2}k_{T}x_{1}\,g\left(x_{1},\,\boldsymbol{p}_{T}-\boldsymbol{k}_{T}\right)\int_{0}^{1}dz\int_{0}^{1}dz'\label{FD1-2}\\
 &  & \times\,\,\,\int\frac{d^{2}r_{1}}{4\pi}\,\int\frac{d^{2}r_{2}}{4\pi}e^{i\left(r_{1}-r_{2}\right)\cdot\boldsymbol{k}_{T}}\,\Psi_{\bar{Q}Q}^{\dagger}\left(r_{2},\,z,\,p_{T}\right)\Psi_{\bar{Q}Q}^{\dagger}\left(r_{1},\,z,\,p_{T}\right)\nonumber \\
 &  & \times N_{M}\left(x_{2}(y);\,\vec{r}_{1},\,\vec{r}_{2}\right)+\left(x_{1}\leftrightarrow x_{2}\right),\nonumber \\
 &  & x_{1,2}\approx\frac{\sqrt{m_{M}^{2}+\langle p_{\perp M}^{2}\rangle}}{\sqrt{s}}e^{\pm y}
\end{eqnarray}
where $y$ and $\boldsymbol{p}_{T}$ are the rapidity and transverse
momenta of the produced heavy meson in the center-of-mass frame of the
colliding protons; $\boldsymbol{k}_{T}$ is the transverse momentum
of heavy quark, $g\left(x_{1},\,\boldsymbol{p}_{T}\right)$ in the
first line of~(\ref{FD1-2}) is the unintegrated gluon PDF; $\Psi_{g\to\bar{Q}Q}(r,\,z)$
is the light-cone wave function of the $\bar{Q}Q$ pair with transverse
separation between quarks $r$ and the light-cone fraction carried
by the quark $z$, and we use for it the standard perturbative expressions~\cite{Rezaeian:2012ji}
\begin{align}
\Psi_{T}^{\dagger}\left(r_{2},\,z,\,Q^{2}\right)\Psi_{T}\left(r_{1},\,z,\,Q^{2}\right) & =\frac{\alpha_{s}N_{c}}{2\pi^{2}}\left\{ \epsilon_{f}^{2}\,K_{1}\left(\epsilon_{f}r_{1}\right)K_{1}\left(\epsilon_{f}r_{2}\right)\left[e^{i\theta_{12}}\,z^{2}+e^{-i\theta_{12}}(1-z)^{2}\right]\right.\\
 & \left.+m_{f}^{2}K_{0}\left(\epsilon_{f}r_{1}\right)K_{0}\left(\epsilon_{f}r_{2}\right)\right\} ,\nonumber \\
\Psi_{L}^{\dagger}\left(r_{2},\,z,\,Q^{2}\right)\Psi_{L}\left(r_{1},\,z,\,Q^{2}\right) & =\frac{\alpha_{s}N_{c}}{2\pi^{2}}\,\left\{ 4Q^{2}z^{2}(1-z)^{2}K_{0}\left(\epsilon_{f}r_{1}\right)K_{0}\left(\epsilon_{f}r_{2}\right)\right\} ,
\end{align}
\begin{equation}
\epsilon_{f}^{2}=z\,(1-z)\,Q^{2}+m_{f}^{2}
\end{equation}
\begin{equation}
\left|\Psi^{(f)}\left(r,\,z,\,Q^{2}\right)\right|^{2}=\left|\Psi_{T}^{(f)}\left(r,\,z,\,Q^{2}\right)\right|^{2}+\left|\Psi_{L}^{(f)}\left(r,\,z,\,Q^{2}\right)\right|^{2}
\end{equation}
The meson production amplitude $N_{M}$ depends on the mechanism of
$Q\bar{Q}$ pair formation. For the case of the two-pomeron fusion,
it is given in leading order by~\cite{Goncalves:2017chx} (see
also Appendix~\ref{sec:Derivation})
\begin{eqnarray}
 &  & N_{M}\left(x,\,\,\vec{r}_{1},\,\vec{r}_{2}\right)=\label{eq:N2}\\
 &  & =-\frac{1}{2}N\left(x,\,\vec{r}_{1}-\vec{r}_{2}\right)-\frac{1}{16}\left[N\left(x,\,\vec{r}_{1}\right)+N\left(x,\,\vec{r}_{2}\right)\right]-\frac{9}{8}N\left(x,\,\bar{z}\left(\vec{r}_{1}-\vec{r}_{2}\right)\right)\nonumber \\
 &  & +\frac{9}{16}\left[N\left(x,\,\bar{z}\vec{r}_{1}-\vec{r}_{2}\right)+N\left(x,\,\bar{z}\vec{r}_{2}-\vec{r}_{1}\right)+N\left(x,\,\bar{z}\vec{r}_{1}\right)+N\left(x,\,\bar{z}\vec{r}_{2}\right)\right].\nonumber 
\end{eqnarray}

For the $p_{T}$-integrated cross-section the gluon uPDF $x_{1}\,g\left(x_{1},\,\boldsymbol{p}_{T}-\boldsymbol{k}_{T}\right)$
must be replaced with the integrated gluon PDF $\,x_{g}G\left(x_{g},\mu_{F}\right),$
which should be taken at the scale $\mu_{F}\,\approx2\,m_{Q}$. In
the LHC kinematics at central rapidities (our principal interest)
this scale significantly exceeds the saturation scale $Q_{s}(x)$,
which justifies the use of two-pomeron approximation. However, in
the kinematics of small-$x_{g}$ (large energies) there are sizable
non-perturbative (nonlinear) corrections to the evolution in the dipole
approach, and therefore in this kinematics the corresponding scale $\mu_{F}$ should
be taken at the saturation momentum $Q_{s}$. In this approach the gluon PDF $x_{1}G\left(x_{1},\,\mu_{F}\right)$
is closely related to the dipole scattering amplitude
$N\left(y,r\right)=\int d^{2}b\,N\left(y,r,b\right)$ as~\cite{KOLEB,THOR}
\begin{equation}
\frac{C_{F}}{2\pi^{2}\bar{\alpha}_{S}}N\left(y,\,\vec{r}\right)=\int\frac{d^{2}k_{T}}{k_{T}^{4}}\phi\left(y,k_{T}\right)\,\Bigg(1-e^{i\vec{k}_{T}\cdot\vec{r}}\Bigg);~~~~x\,G\left(x,\,\mu_{F}\right)=\int_{0}^{\mu_{F}}\frac{d^{2}k_{T}}{k_{T}^{2}}\phi\left(x,\,k_{T}\right),\label{GN1-1}
\end{equation}
where $y=\ln(1/x)$. Eq. (\ref{GN1-1}) can be inverted and it
gives the gluon uPDF in terms of the dipole amplitude, 
\begin{equation}
xG\left(x,\,\mu_{F}\right)\,\,=\,\,\frac{C_{F}\mu_{F}}{2\pi^{2}\bar{\alpha}_{S}}\int d^{2}r\,\frac{J_{1}\left(r\,\mu_{F}\right)}{r}\nabla_{r}^{2}N\left(y,\,\vec{r}\right).\label{GN2-1}
\end{equation}

\begin{figure}
\includegraphics[width=9cm]{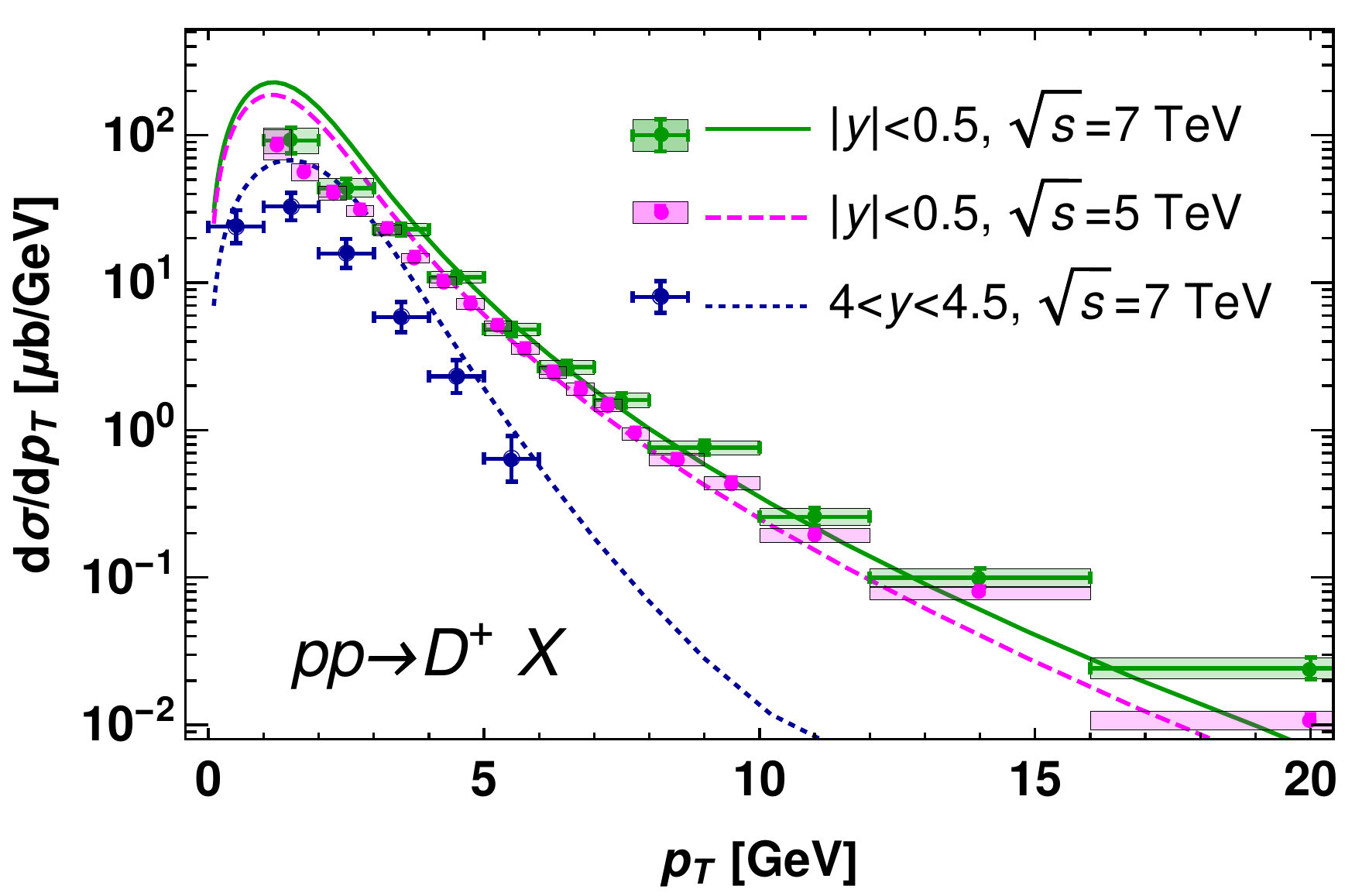}\includegraphics[width=9cm]{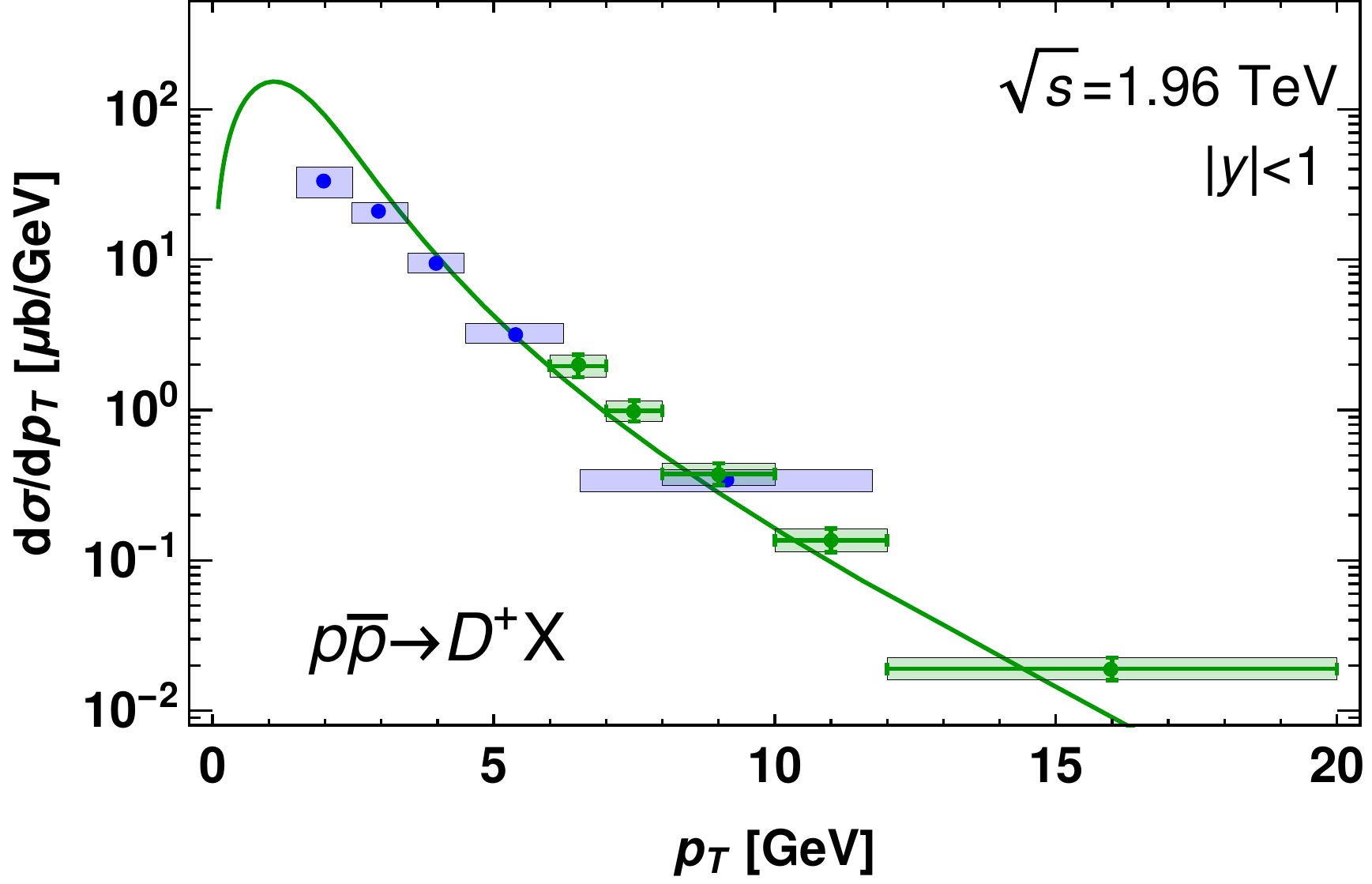}

\caption{\label{fig:pTDependence}The $p_{T}$-dependence of the cross-section
$d\sigma/dp_{T}$ for $D^{+}$-mesons, evaluated with the two-pomeron
fusion mechanism and integrated over the rapidity bin. Left plot:
comparison with data in the LHC kinematics, at central and forward
rapidities. The experimental data are from~\cite{Acharya:2017jgo,Acharya:2019mgn,Aaij:2013mga}.
Right plot: Comparison with experimental data from the Tevatron at central
rapidities. The experimental points are from the CDF and D0 collaborations~\cite{Popov:2017odh,Acosta:2003ax}.
For other mesons the $p_{T}$-dependence has a similar shape, although it
differs by a numerical factor of two (a more detailed comparison with
data can be found in~\cite{Fujii:2013yja,Goncalves:2017chx}).}
\label{Diags_DMesons} 
\end{figure}

\begin{figure}
\includegraphics[width=9cm]{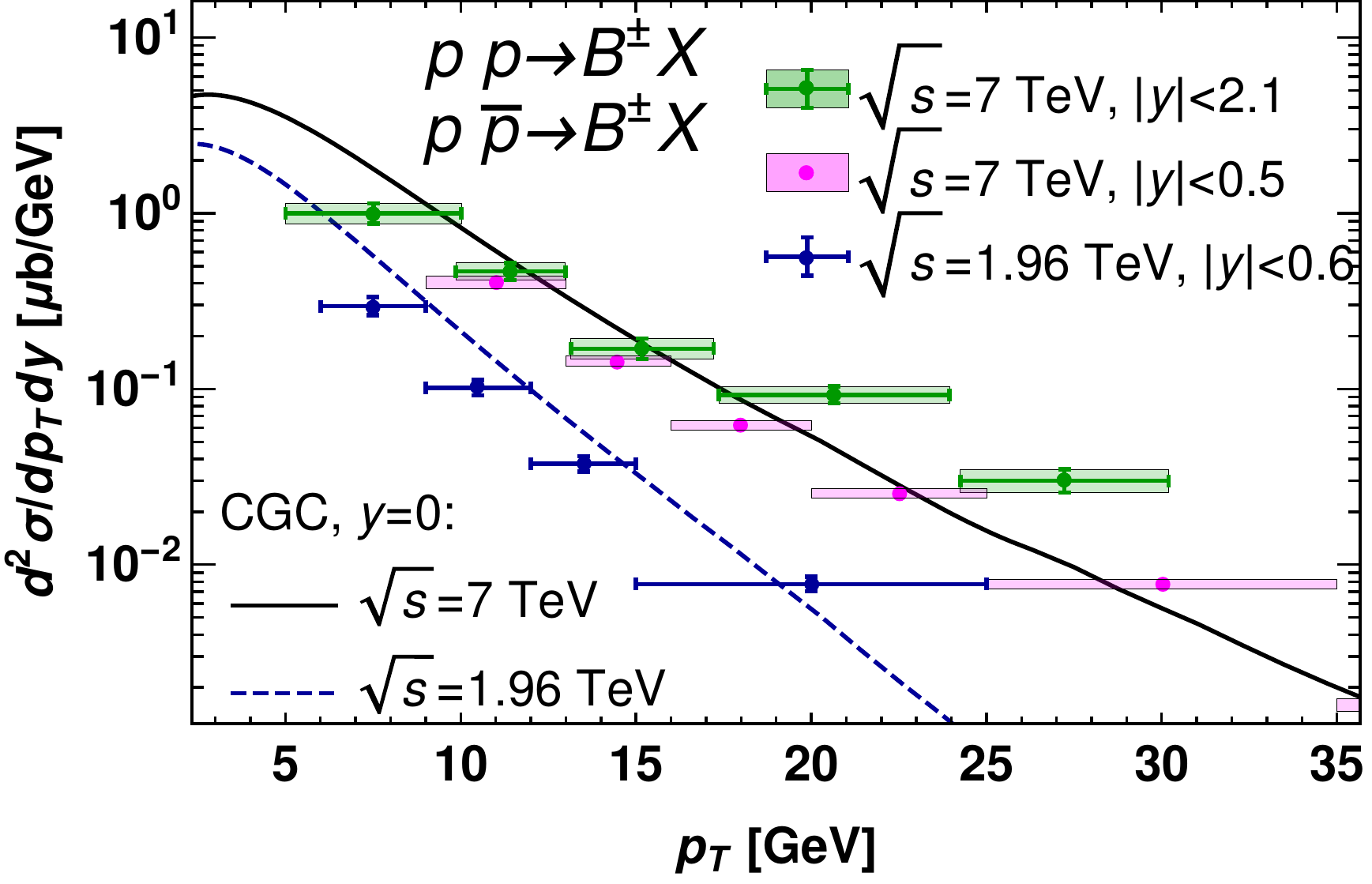}\includegraphics[width=9cm]{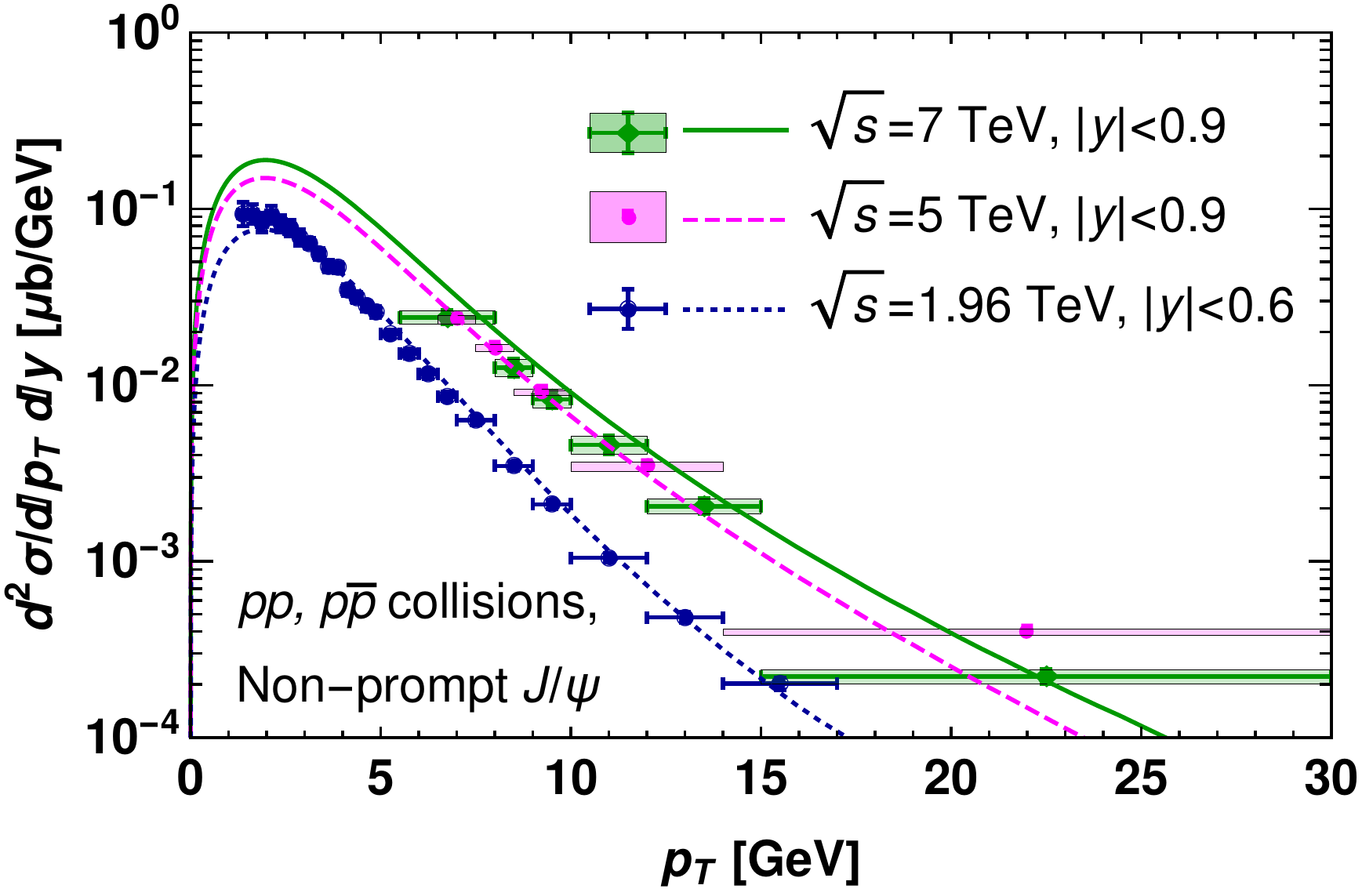}

\caption{\label{fig:pTDependence-1}Left plot: Comparison of experimental and
theoretical results for the $B^{\pm}$-mesons production cross-section
$d\sigma/dy\,dp_{T}$. The experimental data are from CMS~\cite{Khachatryan:2016csy}(``$\sqrt{s}$=7
TeV, $|y|<2.1$`` data points), ATLAS~\cite{ATLAS:2013cia}($\sqrt{s}$=7
TeV, $|y|<0.5$ data points) and CDF~\cite{Abulencia:2006ps} ($\sqrt{s}$=1.96
TeV, $|y|<0.6$ data points). Right plot: The $p_{T}$-dependence
of the cross-section $d\sigma/dy\,dp_{T}$ for non-prompt $J/\psi$.
Comparison with experimental data from CMS~\cite{Sirunyan:2017mzd}
($\sqrt{s}=5$ TeV data) and CDF~\cite{Acosta:2004yw}($\sqrt{s}=1.96$TeV
data) at central rapidities. For $\psi(2S)$ the $p_{T}$-dependence
has a similar shape and differs only by normalization. In both plots,
for some experimentally measured bin-integrated cross-sections $d\sigma/dp_{T}$,
it was converted into $d\sigma/dp_{T}dy$ dividing by the width of the
rapidity bin (this is justified since in the LHC kinematics at central
rapidities $y\approx0$ the cross-section is flat).}
\label{Diags_BMesons} 
\end{figure}

The corresponding unintegrated gluon PDF can be rewritten as~\cite{Kimber:2001sc}
\[
x\,g\left(x,\,k^{2}\right)=\left.\frac{\partial\,}{\partial\mu_{F}^{2}}xG\left(x,\,\mu_{F}\right)\right|_{\mu_{F}^{2}=k^{2}}
\]
which allows to rewrite the result in a symmetric and self-consistent
form, which in turn permits a straightforward generalization of the high-multiplicity
events. Here and in what follows, for our numerical evaluations we
will we use the ``CGC'' parametrization of the dipole cross-section~\cite{RESH}
\begin{align}
N\left(x,\,\vec{\boldsymbol{r}}\right) & =\sigma_{0}\times\left\{ \begin{array}{cc}
N_{0}\,\left(\frac{r\,Q_{s}(x)}{2}\right)^{2\gamma_{{\rm eff}}(r)}, & r\,\le\frac{2}{Q_{s}(x)}\\
1-\exp\left(-\mathcal{A}\,\ln\left(\mathcal{B}r\,Q_{s}\right)\right), & r\,>\frac{2}{Q_{s}(x)}
\end{array}\right.~,\label{eq:CGCDipoleParametrization}\\
 & \mathcal{A}=-\frac{N_{0}^{2}\gamma_{s}^{2}}{\left(1-N_{0}\right)^{2}\ln\left(1-N_{0}\right)},\quad\mathcal{B}=\frac{1}{2}\left(1-N_{0}\right)^{-\frac{1-N_{0}}{N_{0}\gamma_{s}}},\\
 & Q_{s}(x)=\left(\frac{x_{0}}{x}\right)^{\lambda/2},\,\,\gamma_{{\rm eff}}(r)=\gamma_{s}+\frac{1}{\kappa\lambda Y}\ln\left(\frac{2}{r\,Q_{s}(x)}\right),\\
 & \gamma_{s}=0.762,\quad\lambda=0.2319,\quad\sigma_{0}=21.85\,{\rm mb},\quad x_{0}=6.2\times10^{-5}
\end{align}
In Figures~\ref{fig:pTDependence},~\ref{fig:pTDependence-1}
we show the $p_{T}$-dependence for both $D$-meson and $B$-meson production,
as well as for the case of non-prompt $J/\psi$ mesons. We can see that in
the large $p_{T}$ region the two-pomeron mechanism describes very
well all the available data. At small-$p_{T}\lesssim5\,{\rm GeV}$
there are no direct measurements for $B$-mesons, although there are
data for non-prompt $J/\psi$ (from decays of the $B$-mesons), and
we can see that the model describes the data available from Tevatron~\cite{Popov:2017odh,Acosta:2003ax}.
However, for $D$-mesons the agreement is marginal in this kinematics,
and the two-pomeron mechanism systematically overestimates the experimental
data by more than $2\,\sigma$. Such behavior is not related to technical
details of our evaluation (like the choice of the dipole cross-sections
or fragmentation functions) and was also observed by other authors
(see e.g.~\cite{Ma:2018bax,Goncalves:2017chx}). Since this small-$p_{T}$
region gives the dominant contribution to the $p_{T}$-integrated
cross-section, the two-pomeron mechanism will also overestimate this
observable. As we will demonstrate in the next section, the agreement
with data in the small-$p_{T}$ kinematics improves after the inclusion
of the multigluon contributions.

\subsection{Three-pomeron contribution}

\label{subsec:3Pom-1}~

\begin{figure}
\includegraphics[width=9cm]{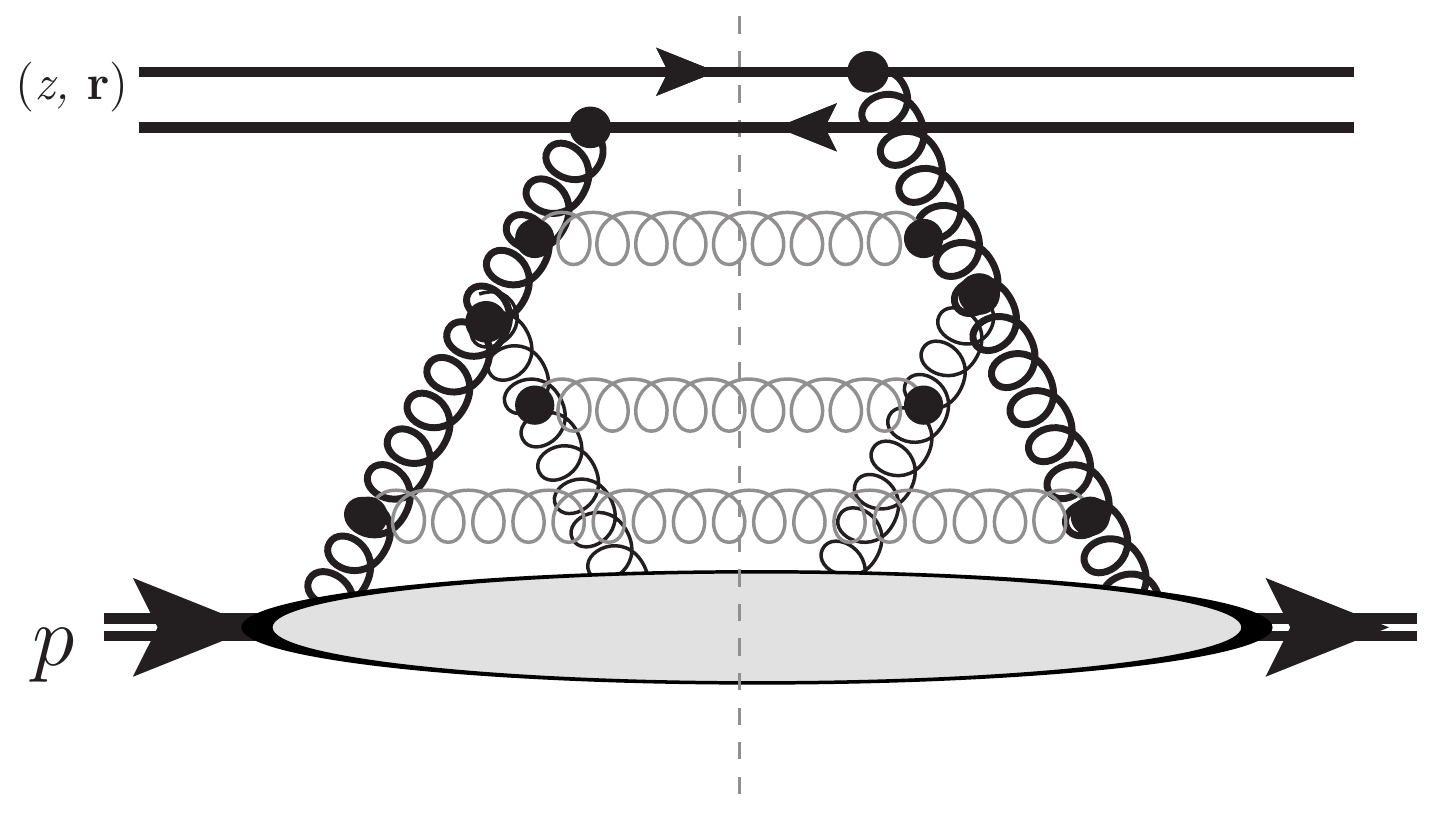}\includegraphics[width=9cm]{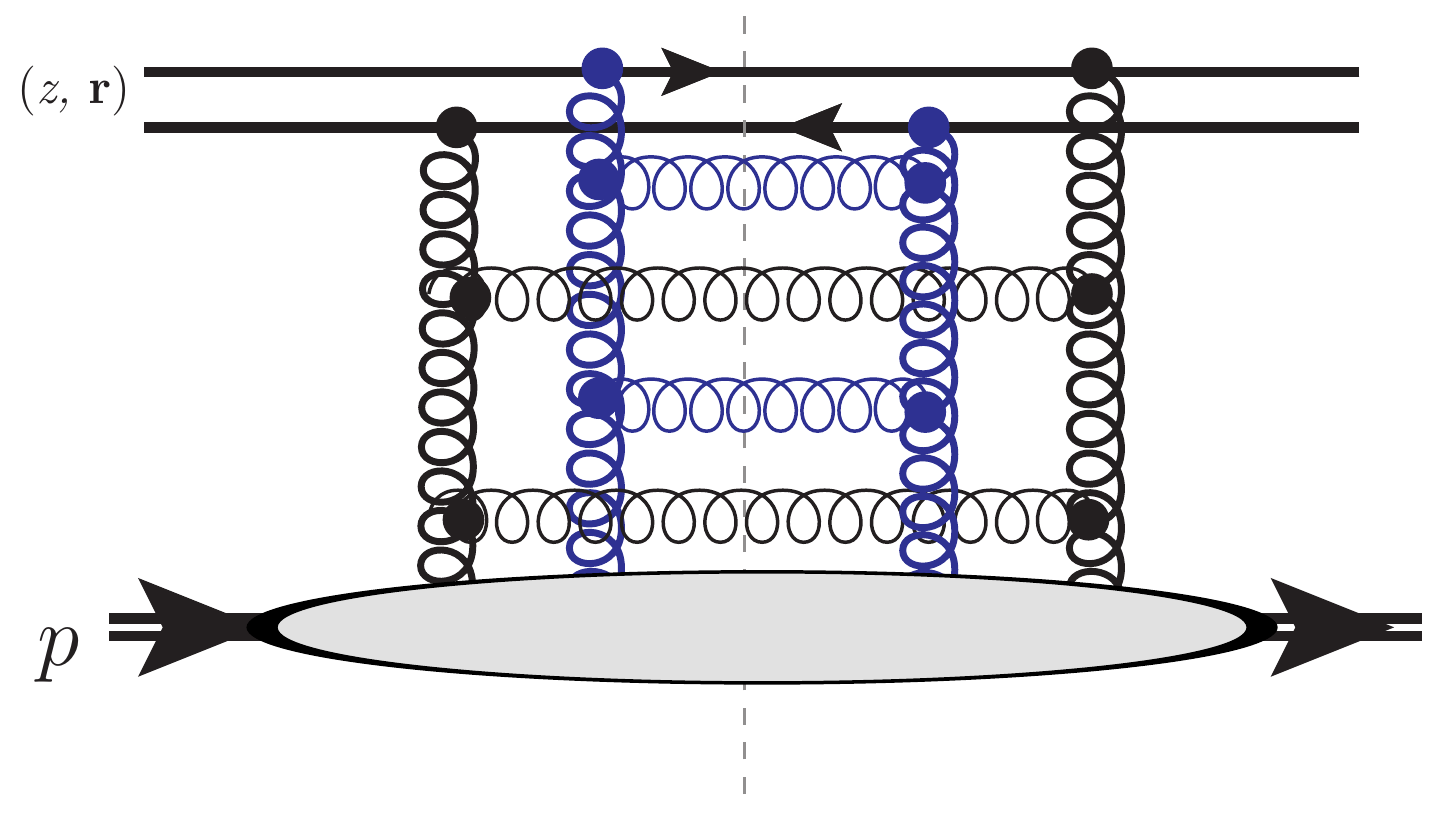}

\caption{\label{fig:DipoleCrossSections}Left plot: A typical fan diagram taken
into account in the CGC parametrization~~\cite{Iancu:2003ge,RESH,Kowalski:2006hc,Watt:2007nr}
of the \emph{color singlet} dipole cross-section $N(z,\,r)$ (resummation
of all possible tree-like topologies is implied).~ Right plot: The
BFKL ladder diagrams resummed in the IP-Sat (\emph{b}-Sat) parametrization~\cite{Kowalski:2003hm,Rezaeian:2012ji}.
In both plots a vertical dashed grey line stands for the unitarity
cut, the blob in the lower part is the hadronic target (proton); two
fermionic lines in the upper part of the blob stand for the dipole
of the transverse size $\boldsymbol{r}$. }
\end{figure}
As we discussed in the introduction, for the $c$-quarks potentially
there could be a sizable contribution from the 3-gluon fusion mechanism.
While usually it is believed that such contributions are suppressed
by $\alpha_{s}\left(m_{Q}\right)$, and in certain cases additionally
by $\Lambda_{{\rm QCD}}^{2}/m_{Q}^{2}$, we have seen from~\cite{Siddikov:2019xvf,Levin:2018qxa}
that potentially such contributions might give a sizable correction
for charmonia production, especially in large-multiplicity
events. In the framework of the dipole model it is usually assumed
that the universal dipole cross-section takes into account all such
contributions. However, in phenomenological parametrizations usually
such contributions either are taken into account with additional simplifying
assumptions or some of the contributions are disregarded. For
example, a widely used phenomenological parametrization ``CGC''
suggested in~\cite{Iancu:2003ge,RESH,Kowalski:2006hc,Watt:2007nr},
was inspired by a solution of the Balitsky-Kovchegov (BK) equation
and effectively resums only fan diagrams shown in the left panel of
Figure~\ref{fig:DipoleCrossSections}. This parametrization does
not take into account the three-pomeron contributions at all. The
competing IP-Sat parametrization~\cite{Kowalski:2003hm,Rezaeian:2012ji},
which was inspired by Glauber-like approach, resums thefor  set of BFKL
ladder diagrams shown in the right panel of Figure~\ref{fig:DipoleCrossSections}.
A central assumption which allows for numerical simplifications is that
the interaction of the BFKL ladder (pomeron) with a dipole of size $r$
is given by $\sim\alpha_{s}\left(\mu^{2}\right)r^{2}x\,g(x)$, which
might work for small color-singlet dipoles, but
in general cases requires a more careful treatment. Although for sufficiently
small dipoles the predictions of both approaches agree with each other,
for the subleading terms the CGC and IP-Sat dipole cross-sections
might differ significantly. For this reason in general we cannot extract
the contribution of the three-pomeron mechanism from~(\ref{FD1-2},~\ref{eq:N2})
and need to evaluate it explicitly. However, we should take into account
that in contrast to $J/\psi$ production at the same order of 
perturbation theory, we may get also interference terms of the leading-order
with subleading order contributions. Since we work in the eikonal approximation,
these diagrams will differ only by a numerical (combinatorial) factor.
Due to these interference contributions the correction is not positively
defined. 

\begin{figure}
\includegraphics[width=18cm]{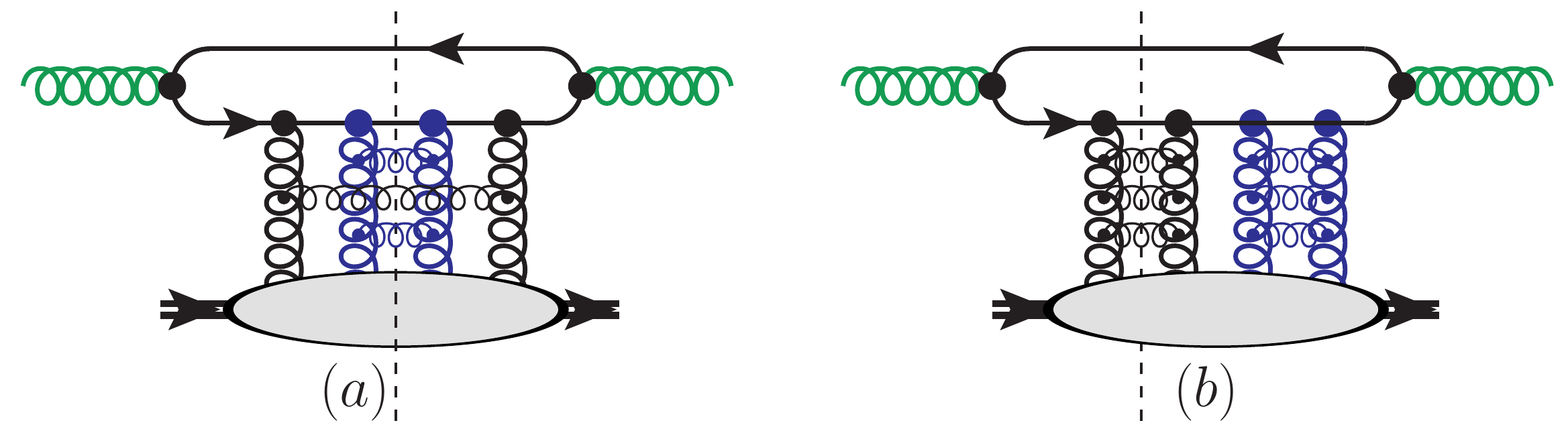}

\caption{\label{fig:NNLOInterference-CutUncut} (color online) The three-pomeron
contributions (diagram ($a$)) contribute at the same order in $\alpha_{s}$
as the interference of LO and NNLO diagrams (diagram ($b$)), and
for this reason the interference terms should be taken into account.
In both plots the vertical dashed line is a unitary cut, lower blob
is a target (proton), and all possible connections of pomerons (thick
wavy lines) to the heavy $Q,\bar{Q}$ quark lines are implied. Note
that in diagram (a) both pomerons are cut, whereas in case of the
interference contribution one of the pomerons is uncut and thus does
not contribute to observed multiplicity enhancement, as explained
in the next Section~\ref{sec:Numer}. }
\end{figure}

As was demonstrated in the Appendix~\ref{sec:Derivation}, for the
three-pomeron contribution we can show that the corresponding cross-section
is given by

\begin{align}
N_{M}^{(3)}\left(x,\,z,\,\vec{\boldsymbol{r}}_{1},\,\vec{\boldsymbol{r}}_{2}\right)\approx & \,\frac{1}{8\sigma_{{\rm eff}}}\left[N_{+}^{2}\left(x,\,z,\,\vec{\boldsymbol{r}}_{1},\,\vec{\boldsymbol{r}}_{2}\right)\left(\frac{3N_{c}^{2}}{8}\right)+N_{-}^{2}\left(x,\,\vec{\boldsymbol{r}}_{1},\,\vec{\boldsymbol{r}}_{2}\right)\left(\frac{\left(43\,N_{c}^{4}-320N_{c}^{2}+720\right)}{72\,N_{c}^{2}}\right)\right.\label{eq:N3Direct}\\
 & \qquad+\left.\frac{\left(N_{c}^{2}-4\right)}{2}N_{+}\left(x,\,z,\,\vec{\boldsymbol{r}}_{1},\,\vec{\boldsymbol{r}}_{2}\right)N_{-}\left(x,\,\vec{\boldsymbol{r}}_{1},\,\vec{\boldsymbol{r}}_{2}\right)\right]\nonumber 
\end{align}
where 
\begin{align}
N_{-}\left(x,\,\vec{\boldsymbol{r}}_{1},\,\vec{\boldsymbol{r}}_{2}\right) & \equiv-\frac{1}{2}\left[N\left(x,\,\vec{\boldsymbol{r}}_{2}-\vec{\boldsymbol{r}}_{1}\right)-N\left(x,\,\vec{\boldsymbol{r}}_{1}\right)-N\left(x,\,\vec{\boldsymbol{r}}_{2}\right)\right]\\
N_{+}\left(x,\,z,\,\vec{\boldsymbol{r}}_{1},\,\vec{\boldsymbol{r}}_{2}\right) & \equiv-\frac{1}{2}\left[N\left(x,\,\vec{\boldsymbol{r}}_{2}-\vec{\boldsymbol{r}}_{1}\right)+N\left(x,\,\vec{\boldsymbol{r}}_{1}\right)+N\left(x,\,\vec{\boldsymbol{r}}_{2}\right)\right]+N\left(x,\,\bar{z}\vec{\boldsymbol{r}}_{1}-\vec{\boldsymbol{r}}_{2}\right)+N\left(x,\,\bar{z}\vec{\boldsymbol{r}}_{1}\right)\\
 & +N\left(x,\,-\bar{z}\vec{\boldsymbol{r}}_{2}+\vec{\boldsymbol{r}}_{1}\right)+N\left(x,\,-\bar{z}\vec{\boldsymbol{r}}_{2}\right)-2N\left(x,\,\bar{z}\left(\vec{\boldsymbol{r}}_{1}-\vec{\boldsymbol{r}}_{2}\right)\right)\nonumber 
\end{align}
and $\sigma_{{\rm eff}}\approx20\,{\rm mb}$ is an effective cross-section
discussed in detail in~\ref{sec:Derivation}. Both functions $N_{\pm}\left(z,\,\vec{\boldsymbol{r}}_{1},\,\vec{\boldsymbol{r}}_{2}\right)$
are invariant with respect to permutation $\boldsymbol{r}_{1}\leftrightarrow\boldsymbol{r}_{2}$.
For the $p_{T}$-integrated cross-sections it is possible to show
that the integration reduces to $\vec{\boldsymbol{r}}_{1}=\vec{\boldsymbol{r}}_{2}=\vec{\boldsymbol{r}}$,
so the cross-sections $N_{\pm}$ simplify to 
\begin{align}
\tilde{N}_{-}\left(x,\,\,\vec{\boldsymbol{r}}\right) & \equiv N_{-}\left(x,\,\vec{\boldsymbol{r}},\,\vec{\boldsymbol{r}}\right)=N\left(x,\,\vec{\boldsymbol{r}}\right)\\
\tilde{N}_{+}\left(x,\,z,\,\vec{\boldsymbol{r}}\right) & \equiv N_{+}\left(x,\,z,\,\vec{\boldsymbol{r}},\,\vec{\boldsymbol{r}}\right)=2N\left(x,\,\bar{z}\vec{\boldsymbol{r}}\right)+2N\left(x,\,z\vec{\boldsymbol{r}}\right)-N\left(x,\,\vec{\boldsymbol{r}}\right)
\end{align}
For the interference term we get in a similar way

\begin{align}
N_{M}^{({\rm int})}\left(x,\,z,\,\vec{\boldsymbol{r}}_{1},\,\vec{\boldsymbol{r}}_{2}\right)= & -\,\frac{3}{16\,\sigma_{{\rm eff}}}\left[2\,N_{+}\left(x,\,z,\,\vec{\boldsymbol{r}}_{1},\,\vec{\boldsymbol{r}}_{2}\right)\tilde{N}_{+}\left(x,\,z,\,\vec{\boldsymbol{r}}_{2}\right)\left(\frac{3N_{c}^{2}}{8}\right)+\right.\label{eq:N3Interf}\\
 & -N_{-}\left(z,\,\vec{\boldsymbol{r}}_{1},\,\vec{\boldsymbol{r}}_{2}\right)\tilde{N}_{-}\left(x,\,\vec{\boldsymbol{r}}_{2}\right)\left(\frac{\left(43\,N_{c}^{4}-320N_{c}^{2}+720\right)}{72\,N_{c}^{2}}\right)+\nonumber \\
 & +\left.\frac{\left(N_{c}^{2}-4\right)}{2}\left(N_{+}\left(z,\,\vec{\boldsymbol{r}}_{1},\,\vec{\boldsymbol{r}}_{2}\right)\tilde{N}_{-}\left(x,\,\vec{\boldsymbol{r}}_{2}\right)+\tilde{N}_{+}\left(x,\,\vec{\boldsymbol{r}}_{2}\right)N_{-}\left(z,\,\vec{\boldsymbol{r}}_{1},\,\vec{\boldsymbol{r}}_{2}\right)\right)\right]\nonumber 
\end{align}
In general the contribution of the interference term~(\ref{eq:N3Interf})
is negative, and larger by magnitude than the direct contribution~(\ref{eq:N3Direct}),
and for this reason the total correction of the three-pomeron mechanism
in general is \emph{negative}. This contribution is strongly suppressed
at large $p_{T}$ because in this kinematics a typical dipole size
$r\sim1/p_{T}$, and the contributions~(\ref{eq:N3Direct},~\ref{eq:N3Interf})
have an additional suppression~$\sim\mathcal{O}\left(r^{2}\right)\sim\mathcal{O}\left(p_{T}^{-2}\right)$
compared to~(\ref{FD1-2},~\ref{eq:N2}). 

\begin{figure}
\includegraphics[width=18cm]{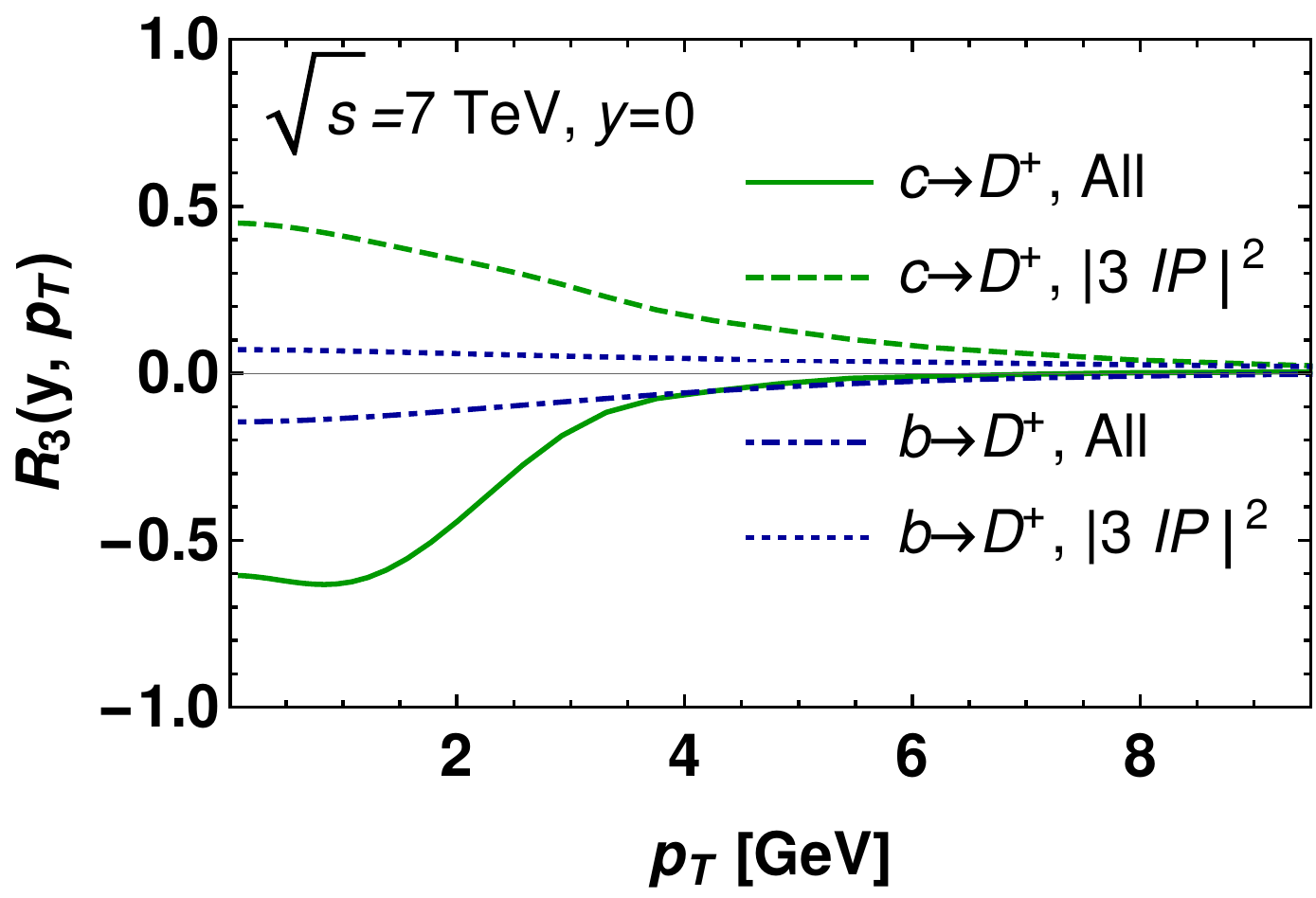}

\caption{\label{fig:pTDependenceRatio}The relative contribution of the 3-gluon
to the 2-gluon mechanism, as defined in~(\ref{eq:R3}). The curves
with labels $"c\to D^{+}"$ and $"b\to D^{+}"$ correspond to prompt
and non-prompt contributions to $D^{+}$-meson production (for other
$D$-mesons the results are similar). The additional label ``$|3\,IP|^{2}$''
in some curves implies that for the contribution of the 3-pomeron
fusion cross-section only the contribution~(\ref{eq:N3Direct}) was
taken into account, whereas for the curves with label ``All'' we
also took into account the contribution of the interference term~(\ref{eq:N3Interf}).
We can see that for $c$-quarks the contribution of the 3-gluon mechanism
in the small-$p_{T}$ kinematics is significant and changes the result
by a factor of two, whereas for $b$-quarks it is just a minor correction
which does not exceed 10\% even for $p_{T}\approx0$. For large $p_{T}$
the relative contribution decreases for all quark flavors, and for
$p_{T}\gtrsim10\,{\rm GeV}$ it becomes negligible. }
\end{figure}

\begin{figure}
\includegraphics[width=9cm]{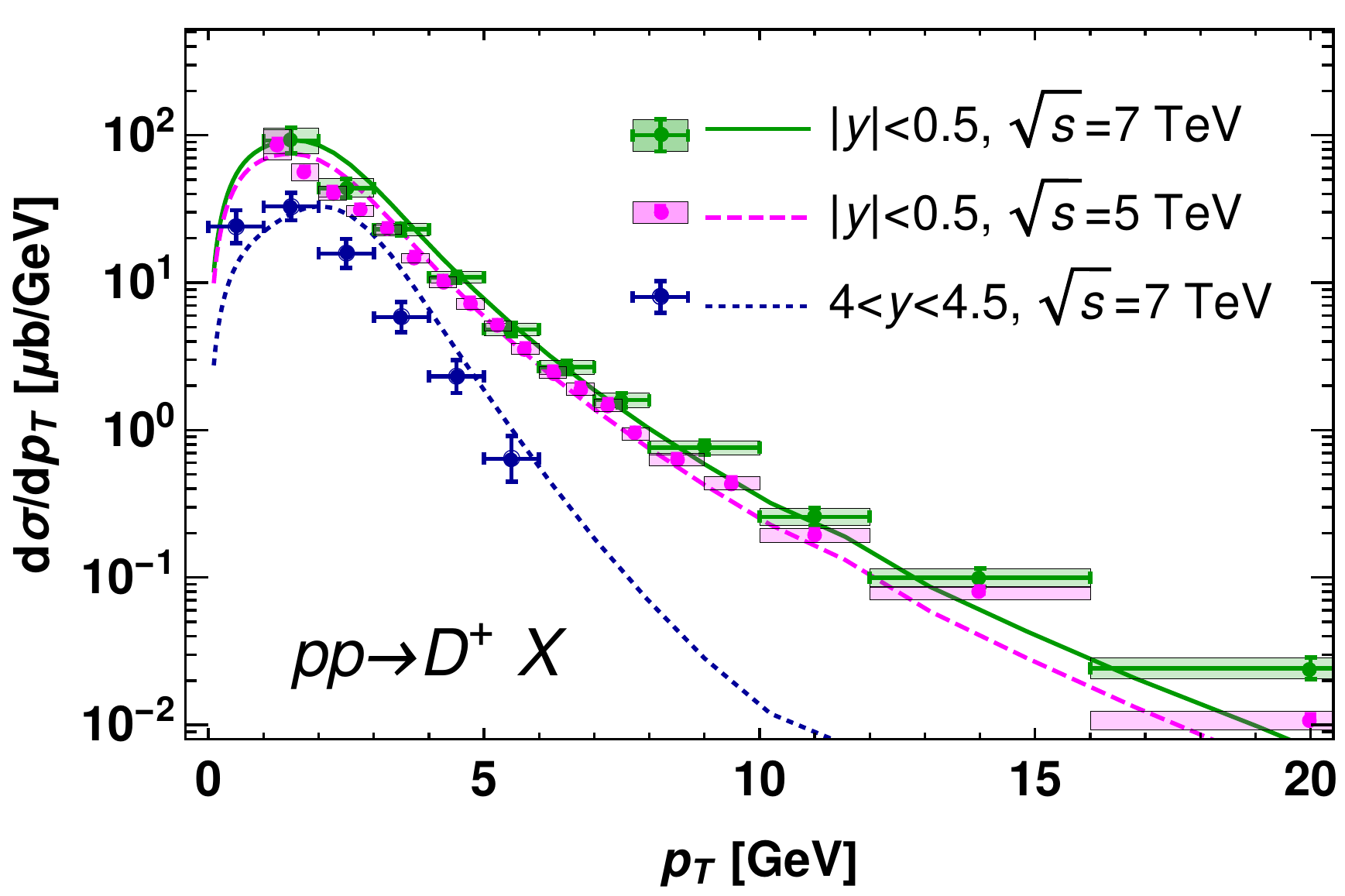}\includegraphics[width=9cm]{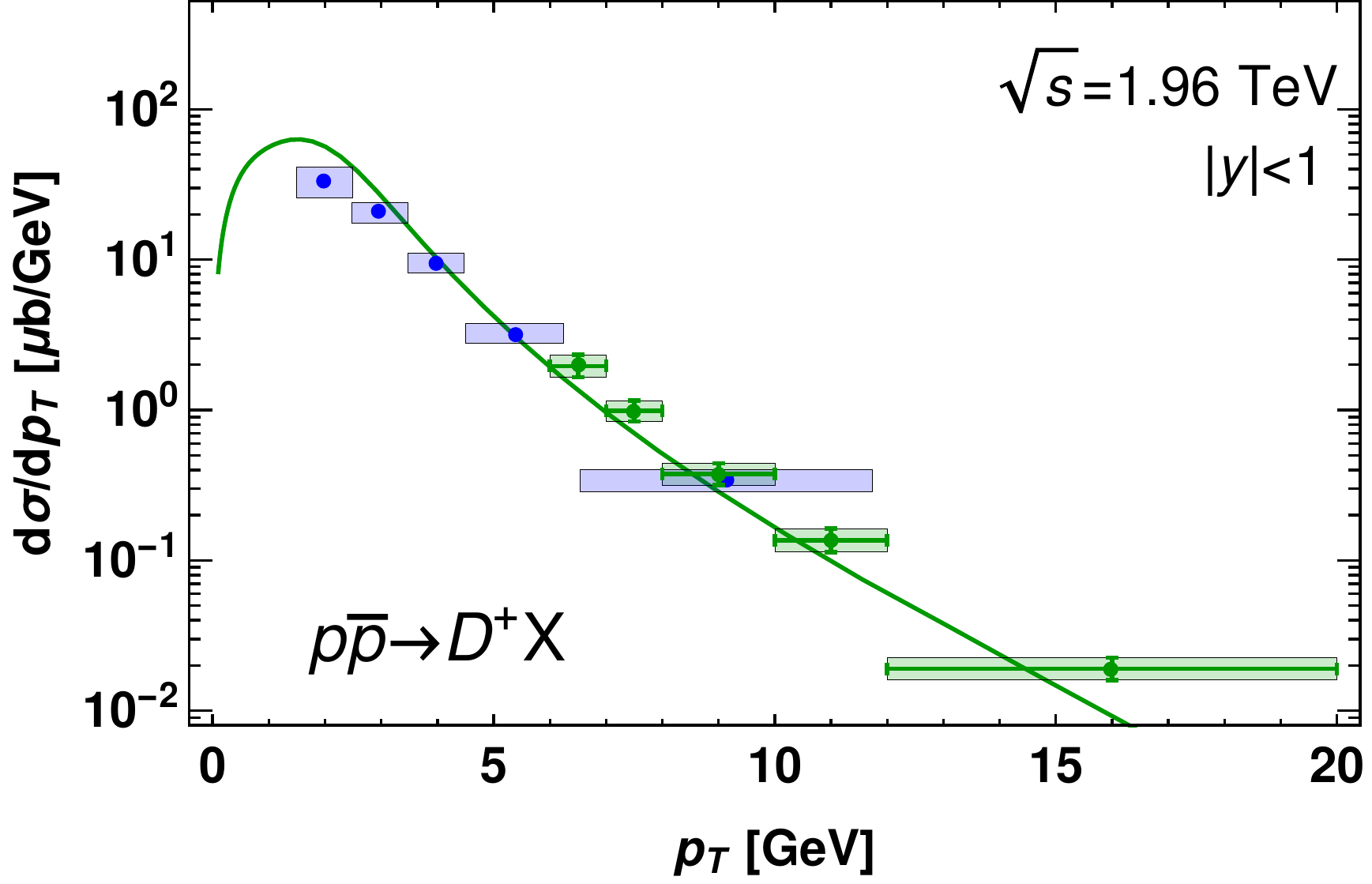}

\caption{\label{fig:pTDependence-2}The $p_{T}$-dependence of the cross-section
$d\sigma/dp_{T}$ for $D^{+}$-mesons, evaluated taking into account both the three-pomeron
and interference contributions. We use the same notations as in the
Figure~(\ref{fig:pTDependence}); the data are integrated over the
rapidity bin. Left plot: comparison with data in the LHC kinematics
at central and forward rapidities. The experimental data are from~\cite{Acharya:2017jgo,Acharya:2019mgn,Aaij:2013mga}.
Right plot: Comparison with experimental data from the Tevatron at central
rapidities. The experimental points are from the CDF and D0 collaborations~\cite{Popov:2017odh,Acosta:2003ax}.
For other mesons the $p_{T}$-dependence has a similar shape, although it
differs by a numerical factor of two (a more detailed comparison with
data might be found in~\cite{Fujii:2013yja,Goncalves:2017chx}).}
\label{Diags_DMesons-1} 
\end{figure}

In order to illustrate the relative size of the three-pomeron mechanism~(\ref{eq:N3Direct})
and the interference term~(\ref{eq:N3Interf}), in Figure~\ref{fig:pTDependenceRatio}
we plotted the ratio of the cross-sections evaluated with three-pomeron
and two-pomeron mechanisms, 
\begin{equation}
R^{(3)}\left(y,\,p_{T}\right)=\frac{d\sigma^{(3)}/dy\,dp_{T}}{d\sigma^{(2)}/dy\,dp_{T}}.\label{eq:R3}
\end{equation}
We can see that for the $c$-quarks at small $p_{T}$ both contributions
might be substantial and constitute up to a factor of two correction.
For the $b$-quarks it does not exceed ten per cent even for $p_{T}\approx0$,
in agreement with heavy quark mass limit expectations. In the large-$p_{T}$
kinematics the relative weight of the three-pomeron contribution is
suppressed and does not exceed a few per cent for $p_{T}\gtrsim10\,{\rm GeV}$.
In Figure~\ref{fig:pTDependence-2} we show the $p_{T}$-dependence
of the cross-section, taking into account both two- and three-pomeron
mechanisms. We can see that in the region of small-$p_{T}$ the agreement
with data is much better than with just the two-pomeron mechanism shown
in~Fig~\ref{fig:pTDependence-1}. For this reason in what follows
we will take into account both the the two- and three-pomeron mechanisms.

\section{Multiplicity dependence}

\label{sec:Numer} 

\subsection{Theoretical framework}

\label{subsec:MultiplicityGeneralities}As was illustrated in the
previous section, the dipole approach ~(\ref{FD1-2}, \ref{eq:N2},
\ref{eq:N3Direct}, \ref{eq:N3Interf}) with a CGC dipole parametrization
provides a very reasonable description of the inclusive $D$- and
$B$-meson production. Nevertheless, the description of the multiplicity dependence
presents more challenges at the conceptual level, because there are
different mechanisms to produce enhanced number of charged particles
$N_{{\rm ch}}$. The probability of multiplicity fluctuations decreases
rapidly as a function of number of produced charged particles $N_{{\rm ch}}$~\cite{Abelev:2012rz},
and for this reason in the study of the multiplicity dependence it is more
common to use a normalized ratio~\cite{Thakur:2018dmp} 
\begin{eqnarray}
 & \frac{dN_{M}/dy}{\langle dN_{M}/dy\rangle}\,\,=\frac{w\left(N_{M}\right)}{\left\langle w\left(N_{M}\right)\right\rangle }\,\frac{\left\langle w\left(N_{{\rm ch}}\right)\right\rangle }{w\left(N_{{\rm ch}}\right)}=\label{eq:NDef}\\
 & =\frac{d\sigma_{M}\left(y,\,\eta,\,\sqrt{s},\,n\right)/dy}{d\sigma_{M}\left(y,\,\eta,\,\sqrt{s},\,\langle n\rangle=1\right)/dy}/\frac{d\sigma_{{\rm ch}}\left(\eta,\,\sqrt{s},\,Q^{2},\,n\right)/d\eta}{d\sigma_{{\rm ch}}\left(\eta,\,\sqrt{s},\,Q^{2},\,\langle n\rangle=1\right)/d\eta}\nonumber 
\end{eqnarray}
where $n=N_{{\rm ch}}/\langle N_{{\rm ch}}\rangle$ is the relative
enhancement of the number of charged particles in the pseudorapidity
window $(\eta-\Delta\eta/2,\,\,\eta+\Delta\eta/2)$; $\langle N_{{\rm ch}}\rangle=\Delta\eta\,dN_{{\rm ch}}/d\eta$
is the average number of charged particles in the pseudorapidity window
$(\eta-\Delta\eta/2,\,\,\eta+\Delta\eta/2)$;$w\left(N_{M}\right)/\left\langle w\left(N_{M}\right)\right\rangle $
and $w\left(N_{{\rm ch}}\right)/\left\langle w\left(N_{{\rm ch}}\right)\right\rangle $
are the self-normalized yields of heavy meson $M$ ($M=D,\,B$) and
charged particles (minimal bias events) in a given multiplicity class;
$d\sigma_{M}(y,\,\sqrt{s},\,n)$ is the production cross-sections
for heavy mesons $M$ with rapidity $y$ and $N_{{\rm ch}}=n\,\langle N_{{\rm ch}}\rangle$
charged particles in the pseudorapidity window $(\eta-\Delta\eta/2,\,\,\eta+\Delta\eta/2)$,
whereas $d\sigma_{{\rm ch}}(y,\,\sqrt{s},\,n)$ is the production
cross-sections for $N_{{\rm ch}}=n\,\langle N_{{\rm ch}}\rangle$
charged particles in the same pseudorapidity window. If the inclusive
cross-section of the process $pp\to M+X$ is proportional to the probability
to produce a meson $M$ in a single $pp$ collision, then the ratio~(\ref{eq:NDef})
gives a \emph{conditional} probability to produce a meson $M$ in
a $pp$ collision in which $N_{{\rm ch}}$ charged particles were
produced. Due to Local Parton-Hadron Duality (LPHD) hypothesis~\cite{LPHD1,LPHD2,LPHD3}
the number of produced charged particles is directly proportional
to the number of partons which stem from the individual pomerons and
thus can be studied using perturbative methods.

In the color dipole approach analyzed in this paper, we expect that
the multiplicity dependence is enhanced due to a larger average number
of particles produced from each pomeron. Nevertheless, we still expect
that each such cascade (``pomeron'') should satisfy the nonlinear
Balitsky-Kovchegov equation, and for this reason we expect that the dipole
amplitude~(\ref{eq:CGCDipoleParametrization}) should maintain its
form, although the value of the saturation scale $Q_{s}$ might be modified.
As was demonstrated in~\cite{KOLEB,KLN,DKLN}, the observed number
of charged multiplicity $dN_{{\rm ch}}/dy$ of soft hadrons in $pp$
collisions is given by the so-called KLN-style formula
\begin{equation}
\frac{dN_{{\rm ch}}}{dy}\,\,=\,\,c\,N_{I\!\!P}\,\frac{Q_{s}^{2}}{\bar{\alpha}_{S}\left(Q_{s}^{2}\right)}\label{MULTQS-2}
\end{equation}
where $c$ is a numerical coefficient, and $N_{I\!\!P}$ is the number
of BK pomerons. Solving algebraic~Eq.(\ref{MULTQS-2}), we could
extract $Q_{s}^{2}$ as a function of $dN_{{\rm ch}}/dy$. Taking
into account that the distribution $dN_{{\rm ch}}/dy$ is almost flat,
we may approximate $n=N_{{\rm ch}}/\langle N_{{\rm ch}}\rangle\approx(dN_{{\rm ch}}/dy)/\langle dN_{{\rm ch}}/dy\rangle$,
so (\ref{MULTQS-2}) allows to express $Q_{s}^{2}$ as a function
of $n$. Frequently in the literature the logarithmic dependence on
$n$, which stems from the running coupling in denominator of~(\ref{MULTQS-2}),
is disregarded, and therefore (\ref{MULTQS-2}) reduces to a simpler linearly
growing dependence on $n$~\cite{KOLEB,KLN,DKLN,Kharzeev:2000ph,Kovchegov:2000hz,LERE,Lappi:2011gu},

\begin{equation}
Q_{s}^{2}\left(x,\,b;\,n\right)\,\,=\,\,n\,Q^{2}\left(x,\,b\right).\label{QSN-1}
\end{equation}
The precision of this assumption was tested in~\cite{Ma:2018bax},
and it was found that a numerical solution of the running coupling
Balitsky-Kovchegov (rcBK) equation differs from the approximate~(\ref{QSN-1})
by less than 10\% in the region of interest ($n\lesssim10$).  Since this
correction is within the precision of current evaluations, in what follows we will use~(\ref{QSN-1}) for our estimates.
While at LHC energies it is expected that the typical values of saturation
scale $Q_{s}\left(x,\,b\right)$ fall into the range 0.5-1 ${\rm GeV}$,
from~(\ref{QSN-1}), we can see that in events with enhanced multiplicity
this parameter might exceed the values of heavy quark mass $m_{Q}$
and lead to an interplay of large-$Q_{s}$ and large-$m_{Q}$ limits.
Thus the study of the high-multiplicity events gives us access to
a new regime which otherwise would require significantly higher energies.

As was shown in~\cite{GLR,MUQI,MV,KOLEB,KLN}, for dilute systems
the saturation scale $Q_{s}$ is closely related to the gluon density
of the target,
\begin{align}
\frac{C_{F}}{2\pi^{2}}\int d^{2}b\,Q_{s}^{2}\left(x,\,\,b,\,n\right)\,\, & =\,x\,G\left(x,\,Q_{s}\right),\label{QS2}\\
C_{F} & \equiv\frac{N_{c}^{2}-1}{2\,N_{c}}.
\end{align}
This qualitative relation just reflects the fact that for dilute system
the saturation scale $Q_{s}^{2}$ is proportional to the density of
partons (gluons), which is described by the gluon density $G$. It
is tempting to extrapolate the relation~(\ref{QS2}) to study the
multiplicity dependence of the gluon density. However, such interpretation
might be useful only for the case when $n$ is not very large, while for
the events with very large multiplicity ($n\gg1$) the concept of
the gluon density becomes quite obscure, since in this case the twist
expansion does not work (it is heavily broken by higher order terms).
The $n$-dependence of~(\ref{QSN-1}) hints to the fact that the contributions
with large number of cut pomerons should be enhanced compared to the
$n=1$ case. Indeed, in the heavy quark mass limit and for not very large
$n$, the typical dipole size in~(\ref{FD1-2}) is given by $\langle r\rangle\sim m_{Q}^{-1}$,
so from the structure of~(\ref{eq:CGCDipoleParametrization}) we
can see that this enhancement is given by a factor $\sim n^{\gamma_{{\rm eff}}}$.
However, in the deeply saturated regime ($n\gg1$), when $Q_{s}^{2}\left(x,\,b;\,n\right)\gtrsim m_{Q}^{2}$,
the typical dipole size is controlled by the saturation scale and
thus the $n$-dependence should be the same for all multipomeron contributions.
We would also like to mention that the uncut pomerons do not contribute
to the observed enhancement of charged particles and thus should not
be taken into account in the multiplicity evaluation.

To conclude, the suggested mechanism introduces a dependence on multiplicity
of soft produced particles, and it is quite different from other approaches
such as the percolation approach~\cite{PER} or the modification of the
slope of the elastic amplitude~\cite{Kopeliovich:2013yfa}.  Moreover, it can
be applied both to the production of soft and hard particles. In the
following subsection we will use this approach for analyzing the multiplicity dependence
of quarkonia production.

\subsection{Phenomenological estimates}

\begin{figure*}
\includegraphics[width=9cm]{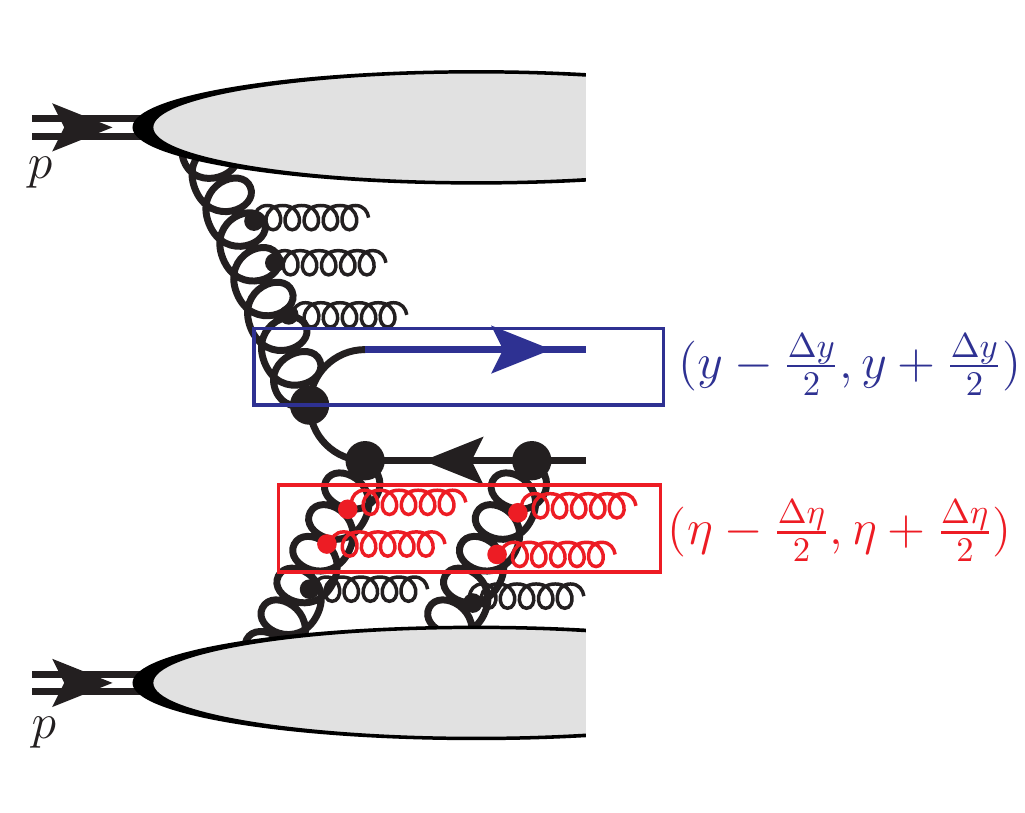}\includegraphics[width=9cm]{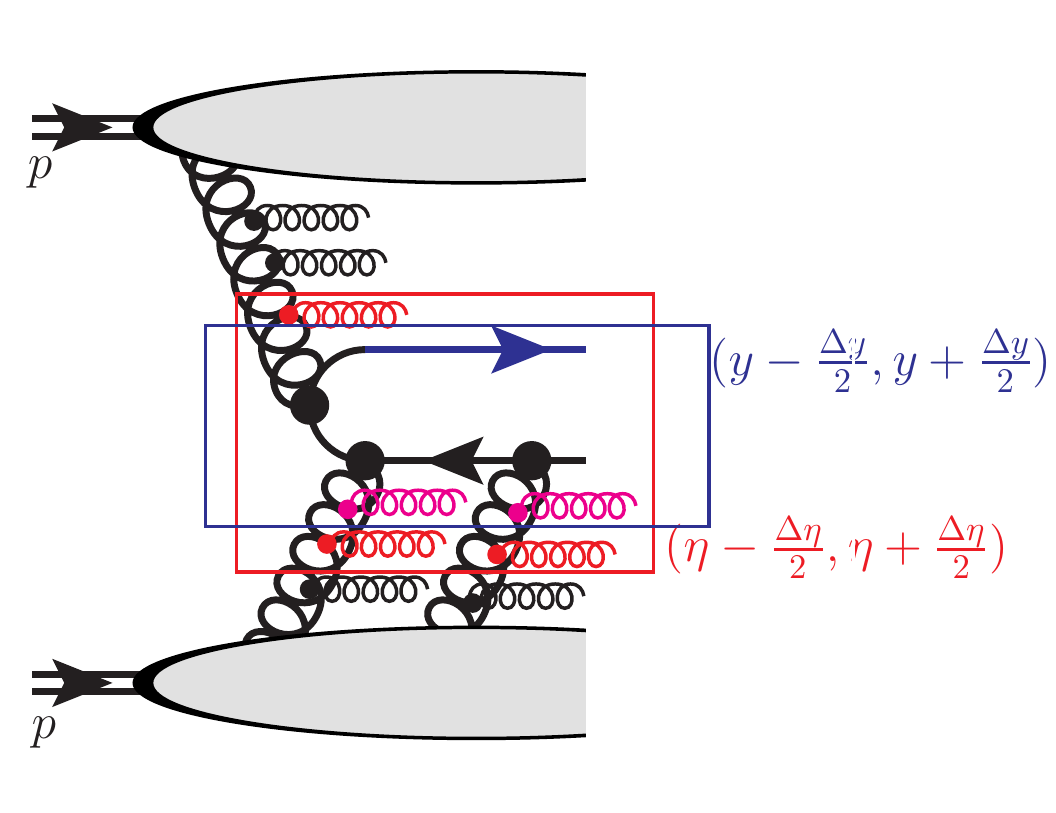}

\caption{(color online) Demonstration that only some pomerons should be taken
into account in the evaluation of enhanced multiplicity. For definiteness we 
consider only the three-pomeron fusion mechanism and for simplicity disregard
the heavy quark fragmentation (convolution with fragmentation
function), which slightly changes the rapidity. Left plot: When the
rapidity bin used for the collection of heavy mesons (blue box) does not
overlap with the bin used for the collection of charged particles (red
box), the elevated multiplicity should be unambiguously attributed
to (shared between) the two lower pomerons. Right plot: Situation
when the bins partially overlap. For the partons in the intersection
region (magenta color) the assignment to upper or lower bins depends
on the position of heavy meson inside the bin~($y-\Delta y/2,\,y+\Delta y/2$).
In the final result we should average over all possible rapidities
of heavy meson inside the bin. \label{fig:DiagMultiplicityDistribution}}
\end{figure*}

In a typical experimental setup the detector used to collect charged
particles $N_{{\rm ch}}$ usually covers some small rapidity bin $\Delta\eta$,
in which a relative enhancement of multiplicity $n=N_{{\rm ch}}/\langle N_{{\rm ch}}\rangle$
is observed. This enhancement is a result of superpositions of increased
multiplicities from individual pomerons. Since each pomeron hadronizes
independently, we may expect that we should apply the modification
of the dipole cross-section discussed in the previous section to each
of the pomerons, modifying the corresponding dipole amplitude. However,
the number of pomerons which can participate in the observed multiplicity
enhancement depends crucially on the details of how the experiment is
done, as explained in Figure~\ref{fig:DiagMultiplicityDistribution}. 

For the simplest case when the bins used for the collection of mesons
and charged particles are well-separated in rapidity (left panel in
Figure~\ref{fig:DiagMultiplicityDistribution}), it is clear
that all charged particles can stem only from cut pomerons at a
given rapidity (lower pomerons in Figure~\ref{fig:DiagMultiplicityDistribution}),
so the cross-section to produce heavy meson $M$ and $N$ particles
can be found as a \emph{mathematical expectation,} convoluting the
probability of a given partition ($N_{1},...N_{k}$) with the value
of the corresponding cross-section, \emph{viz}.:
\begin{align}
\frac{d\sigma_{pp\to\bar{Q}_{i}Q_{i}+X}\left(y,\,\sqrt{s},\,\,n\right)}{dy\,d^{2}p_{T}} & =\left(\prod_{k}\sum_{N_{k}=0}^{N}P\left(N_{k},\,\left\langle N_{k}\right\rangle \right)\right)\delta_{N,\,\sum_{k}N_{k}}\times\label{eq:ModDiff}\\
 & \times\frac{d\sigma_{pp\to\bar{Q}_{i}Q_{i}+X}\left(y,\,\sqrt{s},\,\,n_{1}...n_{k}\right)}{dy\,d^{2}p_{T}},\nonumber \\
 & \quad n_{k}\equiv\frac{N_{k}}{\langle N_{k}\rangle}\nonumber 
\end{align}
In~(\ref{eq:ModDiff}) we sum over all possible partitions of the
number of observed charged particles $N\equiv n\,\Delta\eta=N_{1}+N_{2}+...+N_{k}$.
The additional arguments $n_{1}...n_{k}$ in the arguments of the
cross-section in the second line of~(\ref{eq:ModDiff}) imply a modification
of the saturation scale in a dipole cross-section of individual BK
pomerons by factors $n_{1}...n_{k}$, as described in the previous
section. The function $P\left(N_{k},\,\left\langle N_{k}\right\rangle \right)$
in the integrand is the probability that a single pomeron has a given
multiplicity fluctuation with a mean value $\langle N\rangle$. This
distribution should satisfy a convolution identity

\begin{eqnarray}
 &  & \sum_{N_{1}}\,P\left(N_{1},\,\left\langle N_{1}\right\rangle \right)P\left(N-N_{1},\,\left\langle N_{2}\right\rangle \right)=P\left(N,\,\left\langle N_{1}\right\rangle +\left\langle N_{2}\right\rangle \right).\label{eq:conv-1}
\end{eqnarray}
which implies that a contribution of a pomeron to the observed number
of charged particles equals the sum of all possible contributions of its
parts. The exact evaluation of~(\ref{eq:ModDiff}) requires knowledge
of the function $P\left(N,\,\left\langle N\right\rangle \right)$, which
is an essentially nonperturbative object, studied
in the literature in the context of different models (see e.g.~\cite{Levin:1993te}
for details), and which has been suggested that it
might be described by the Poisson distribution. Fortunately, for phenomenological
estimates we may minimize the sensitivity to the choice of the model for
$P\left(N,\,\left\langle N\right\rangle \right)$, taking into account
the following facts:
\begin{itemize}
\item The dependence on $n_{k}$, which stems from the cross-section in the
second line of~(\ref{eq:ModDiff}), is very mild and saturates (becomes
constant) in the region of large $n$. In contrast, the functions
$P\left(N_{k},\,\left\langle N_{k}\right\rangle \right)$ decrease
exponentially for $n_{k}\equiv N_{k}/\langle N_{k}\rangle\gg1$, and for
this reason in the evaluation of~(\ref{eq:ModDiff}) we may replace each
$n_{k}$ with some average value $\langle n_{k}\rangle$ (which in
general depends on $n$).
\item Since all active pomerons have the same average number of particles
$\langle N_{k}\rangle$, taking into account a very mild dependence
of the cross-section on $n_{k}$, we may apply iteratively~(\ref{eq:conv-1})
and rewrite~(\ref{eq:ModDiff}) as
\begin{align}
\frac{d\sigma_{pp\to\bar{Q}_{i}Q_{i}+X}\left(y,\,\sqrt{s},\,\,n\right)}{dy\,d^{2}p_{T}} & =P\left(N,\,\left\langle N\right\rangle \right)\frac{d\sigma_{pp\to\bar{Q}_{i}Q_{i}+X}\left(y,\,\sqrt{s},\,\,\langle n_{1}\rangle...\langle n_{k}\rangle\right)}{dy\,d^{2}p_{T}}\label{eq:ModDiff-1}\\
\left\langle n_{i}\right\rangle  & \equiv n/k,\quad i=1,...,k.\nonumber 
\end{align}
\end{itemize}
Thus effectively we come to the conclusion that the observed multiplicity
is shared equally between all the pomerons at a given rapidity window.
The convolution of the distributions $P\left(n_{i}\right)$ just cancels
in the ratio~(\ref{eq:NDef}), so the latter can be rewritten as

\begin{eqnarray}
 & \frac{dN_{M}/dy}{\langle dN_{M}/dy\rangle}\,\,=\frac{d\tilde{\sigma}_{pp\to Q\bar{Q}+X}\left(y,\,\eta,\,Q^{2},\,n\right)/dy}{d\tilde{\sigma}_{pp\to Q\bar{Q}+X}\left(Y\,\eta,,\,Q^{2},\,\langle n\rangle=1\right)/dy}\label{eq:NDef-2}
\end{eqnarray}
where we use notation $d\tilde{\sigma}$ instead of $d\sigma$ for
the cross-section, to emphasize that we took out the normalization to probability
distribution of charged particles (factor $P(N,\langle N\rangle)$)
and share multiplicity equally between all pomerons in a given rapidity
window. 

The situation becomes more complicated for the setup when the heavy
meson and charged particles bins partially overlap, as shown in the
right panel of Figure~\ref{fig:DiagMultiplicityDistribution}.
In this case in the intersection region the enhanced multiplicity
is due to either the upper or to the lower cut pomerons, depending
on the position of the heavy quark inside the bin. For the sake of
simplicity we will consider the case of complete overlap of both bins,
which is realized in the case of ALICE measurements at central rapidities.
In this case we have to average over the rapidity of heavy quark in the
relation~(\ref{eq:ModDiff}), taking into account that for the upper
pomerons $\langle N_{k}\rangle\sim(dN/dy)\,(y-y_{\min})$, whereas
for the lower pomerons~$\langle N_{k}\rangle\sim(dN/dy)\,\left(y_{\max}-y\right)$.
Following the assumptions formulated after~Eq.~(\ref{eq:conv-1}),
we can see that instead of~(\ref{eq:NDef-2}) we should use
\begin{eqnarray}
 & \frac{dN_{M}/dy}{\langle dN_{M}/dy\rangle}\,\,=\frac{\int_{y_{\min}}^{y_{\max}}dy\,d\tilde{\sigma}_{pp\to Q\bar{Q}+X}\left(y,\,\eta,\,Q^{2},\,n\right)/dy}{\Delta y\,d\tilde{\sigma}_{pp\to Q\bar{Q}+X}\left(Y\,\eta,,\,Q^{2},\,\langle n\rangle=1\right)/dy}\label{eq:NDef-3}
\end{eqnarray}
with values $n_{k}=n\left(y-y_{{\rm min}}\right)/\left(k_{{\rm up}}\,\Delta y\right)$
and $n_{k}=n\left(y_{\max}-y\right)/\left(k_{{\rm down}}\,\Delta y\right)$
where $k_{{\rm up}},\,k_{{\rm down}}$ is the number of pomerons in
the the upper and lower parts of the diagram. 

\begin{figure*}
\includegraphics[width=9cm]{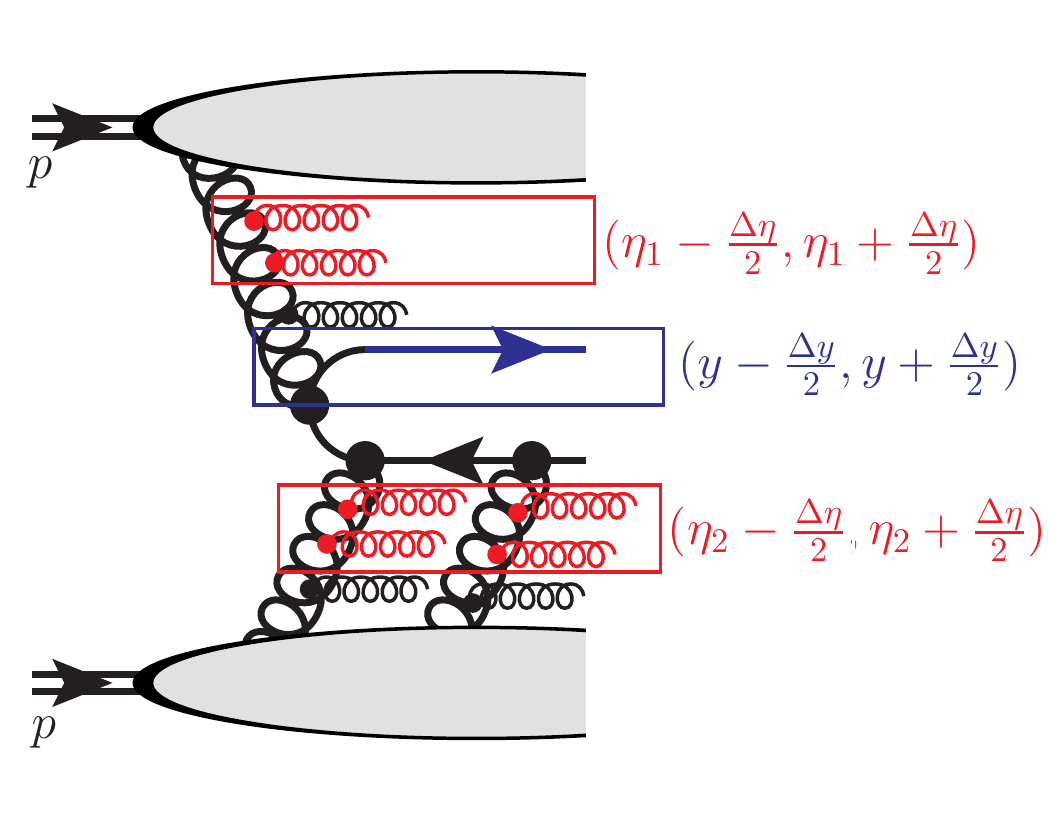}

\caption{(color online) Case of ``minimal bias'' measurement done by the V0 detector
at ALICE. The charged particles are collected in both backward and forward
directions, and only their \emph{sum} is used to measure $N_{{\rm ch}}$.
For the sake of definiteness we consider only the three-pomeron fusion mechanism
and for simplicity disregard the fragmentation of heavy quark (convolution
with fragmentation function), which slightly smears the distribution
over rapidity. The rapidity bin used for the collection of heavy mesons
(blue box) does not overlap with the bins used for the collection of charged
particles (red boxes); however, the elevated multiplicity cannot be
unambiguously attributed to be (shared between) the upper or lower pomerons.
\label{fig:DiagMultiplicityDistribution-V0}}
\end{figure*}

Finally, we would like to stop briefly on the case of the so-called
minimum bias configuration studied by the $V0$ detector at ALICE. In
this case the total charge is accumulated in two rapidity bins, in
very forward and very backward directions, as seen from Figure~\ref{fig:DiagMultiplicityDistribution-V0}.
In this configuration we do not have overlap of $y$- and $\eta$-bins,
yet still we cannot assign the enhanced multiplicity either to upper
or to lower pomerons. Instead of this, we should use a sum of all possible
partitions of the total number of charged particles. The corresponding
cross-section is given by~(\ref{eq:NDef-2}), although we should use $n_{k}=n/k_{{\rm up}}$
and $n_{k}=n/k_{{\rm down}}$, where $k_{{\rm up}},\,k_{{\rm down}}$
is the number of pomerons in the upper and lower parts of the diagram.

\begin{figure}
\includegraphics[width=18cm]{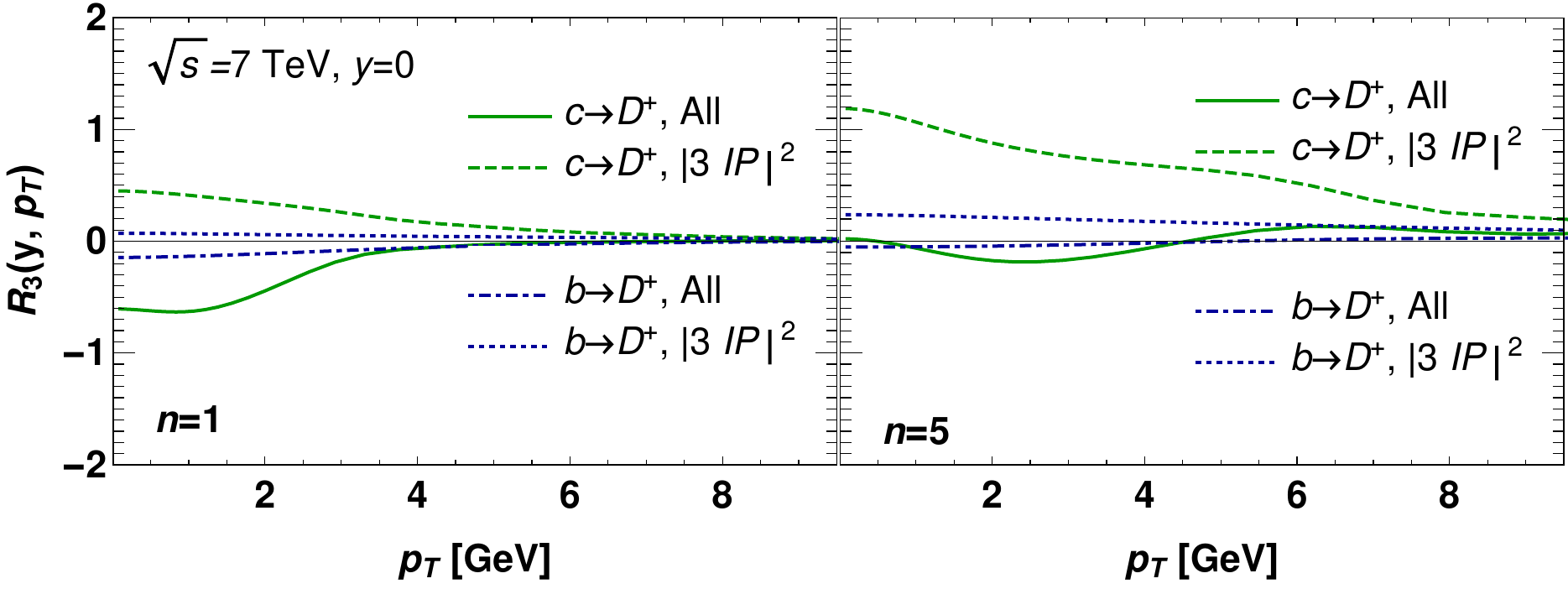}

\caption{\label{fig:pTDependenceRatio-1}Left plot: Relative contribution of
the 3-pomeron to the 2-pomeron mechanism, as defined in~(\ref{eq:R3}).
The curves with labels $"c\to D^{+}"$ and $"b\to D^{+}"$ correspond
to prompt and non-prompt contributions to $D^{+}$-meson production
(for other $D$-mesons the result is similar). The additional label
``$|3\,IP|^{2}$'' in some curves implies that for the contribution
of the 3-pomeron fusion cross-section only the contribution~(\ref{eq:N3Direct})
was taken into account, whereas for the curves with label ``All''
we also took into account the contribution of the interference term~(\ref{eq:N3Interf}).
We can see that for $c$-quarks the contribution of the 3-pomeron
mechanism is substantial and constitutes up to 50\% of the total
result at small $p_{T}$, whereas for $b$-quarks it does not exceed
10\% even for $p_{T}\approx0$. For large $p_{T}$ the relative contribution
decreases both for the $c$- and $b$-quarks, and for $p_{T}\gtrsim10\,{\rm GeV}$
becomes negligible. Right plot: the same evaluation for high-multiplicity
events with $n=5$. The total result (sum of direct and interference
terms) \emph{decreases} as a function of multiplicity and nearly vanishes
for the high-multiplicity events near $n\approx5$, as explained in
the text.}
\end{figure}

We would like to start the discussion of the numerical results from estimates
of the role of the three-pomeron mechanism. In Figure~\ref{fig:pTDependenceRatio-1}
we plotted the ratio of the three-pomeron and two-pomeron contributions~(\ref{eq:R3}),
which was discussed in Section~\ref{subsec:3Pom-1}. We can see that
in the large-multiplicity events the relative contribution of the
three-pomeron mechanism is sizable even for $n=1$. As a function
of multiplicity this ratio grows, and starting from $n\approx5$
for $D$-mesons it becomes a dominant contribution, as can be seen
from the right panel of Figure~\ref{fig:pTDependenceRatio-1}.
Such increase can be understood from a comparison of the structures
of~(\ref{eq:N2},\ref{eq:N3Direct}): in the heavy quark mass limit
the size of the dipole is given by $\langle r_{Q}\rangle\sim1/{\rm \max}\left(m_{Q},\,p_{T}\right)$,
and as we explained in Section~\ref{subsec:MultiplicityGeneralities},
in this regime each cut pomeron yields an additional factor $\sim n^{\gamma_{{\rm eff}}}$.
However, such grow cannot continue up to infinity. When the saturation
scale $Q_{s}$ becomes significantly larger than all the other scales,
the relevant dipole size is given by $\langle r_{Q}\rangle\sim1/Q_{s}$
and the system \emph{saturates}, \emph{i.e}, having a very mild dependence
on $n$. The interference term~(\ref{eq:N3Interf}) has the same
number of cut pomerons as the two-pomeron mechanism, and for this reason we
expect that the ratio~(\ref{eq:R3}) should be largely independent on $n$.
Since at $n=1$ the contribution of the interference term is larger
than (\ref{eq:N3Direct}) and has opposite sign, their sum \emph{decreases}
by absolute value in the range $1\lesssim n\lesssim5$, and changes
sign near $n\approx5$. For this reason a combined contribution of
the three-pomeron mechanisms~(\ref{eq:N3Direct},\ref{eq:N3Interf})
does not affect the multiplicity dependence significantly in the multiplicity
range studied in~\cite{Adam:2015ota}. We also may notice that due
to the increase of the saturation scale $Q_{s}^{2}$, the transition to the
large-$p_{T}$ regime happens at significantly larger $p_{T}$. 

\begin{figure}
\includegraphics[width=18cm]{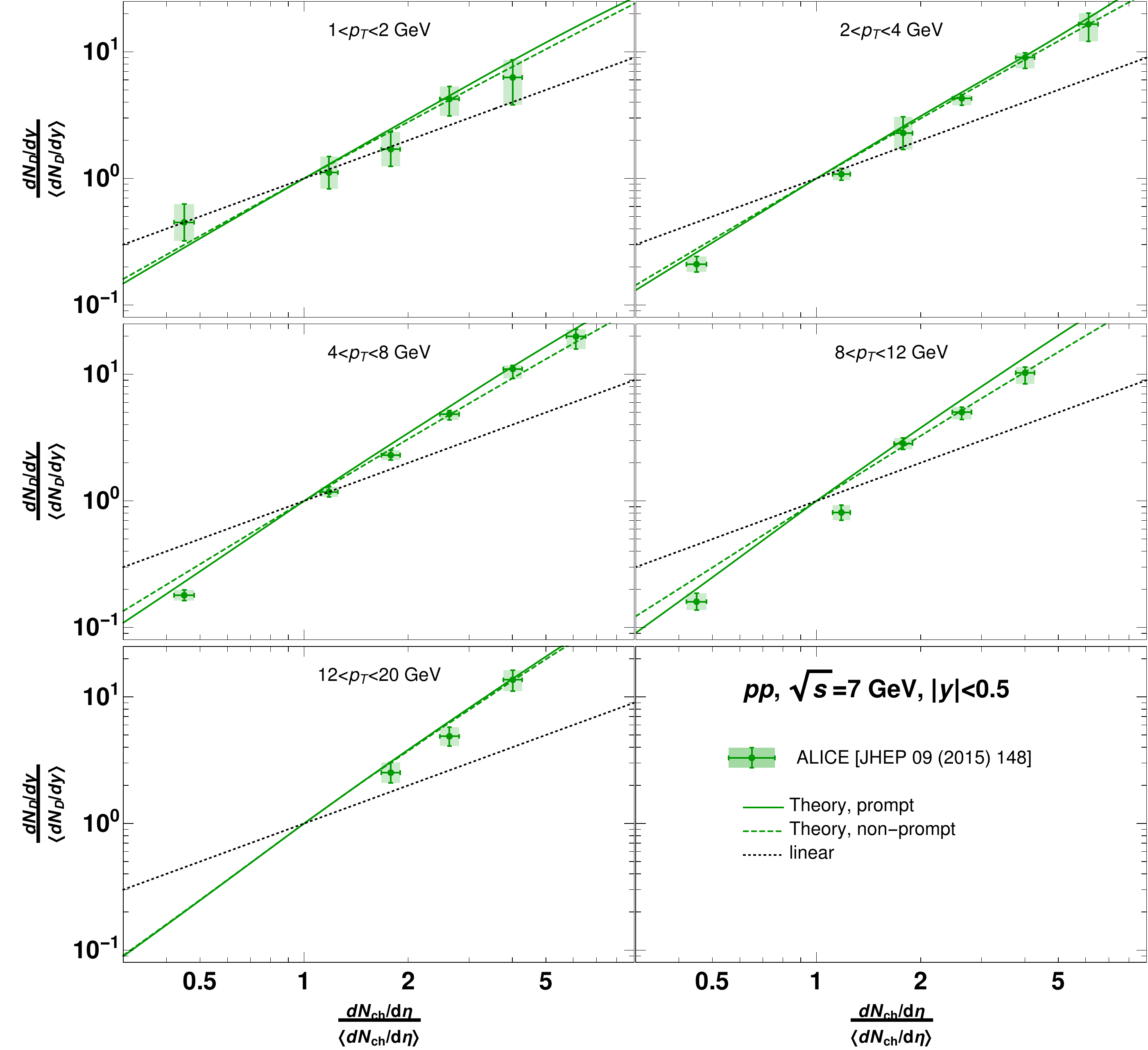}

\caption{\label{DiagsMultiplicityDMesons}Theoretical multiplicity dependence
of the prompt and non-prompt mechanisms of the $D$-meson production
in different bins in $p_{T}$ at central rapidities. The experimental
data are from ALICE~\cite{Adam:2015ota} and correspond to \emph{prompt}
production mechanism for averaged contribution of $D^{0},\,D^{+}$
and $D^{*+}$ mesons. For the sake of reference we have also shown
a ~dotted line which corresponds to a linear dependence. The charged
particles and $D$-mesons are collected in rapidity window $|y|<0.5$,
$|\eta|<1$.}
 
\end{figure}

Currently the data on the multiplicity dependence of open charm mesons
are available from the ALICE experiment~\cite{Adam:2015ota}. As
we can see from Figure~\ref{DiagsMultiplicityDMesons}, our results
can perfectly describe the available data on $D$-meson production.
For the sake of definiteness we make the comparison with experimentally
available averaged contribution of the $D^{+},\,D^{0}$ and $D^{*0}$
mesons. In the same figure we have shown the contribution of the non-prompt
mechanism (dashed lines). As expected, this contribution has the same
dependence on $n$, though numerically its relative size in the inclusive
cross-section varies depending on the kinematics (values of $p_{T}$).

\begin{figure}
\includegraphics[width=12cm]{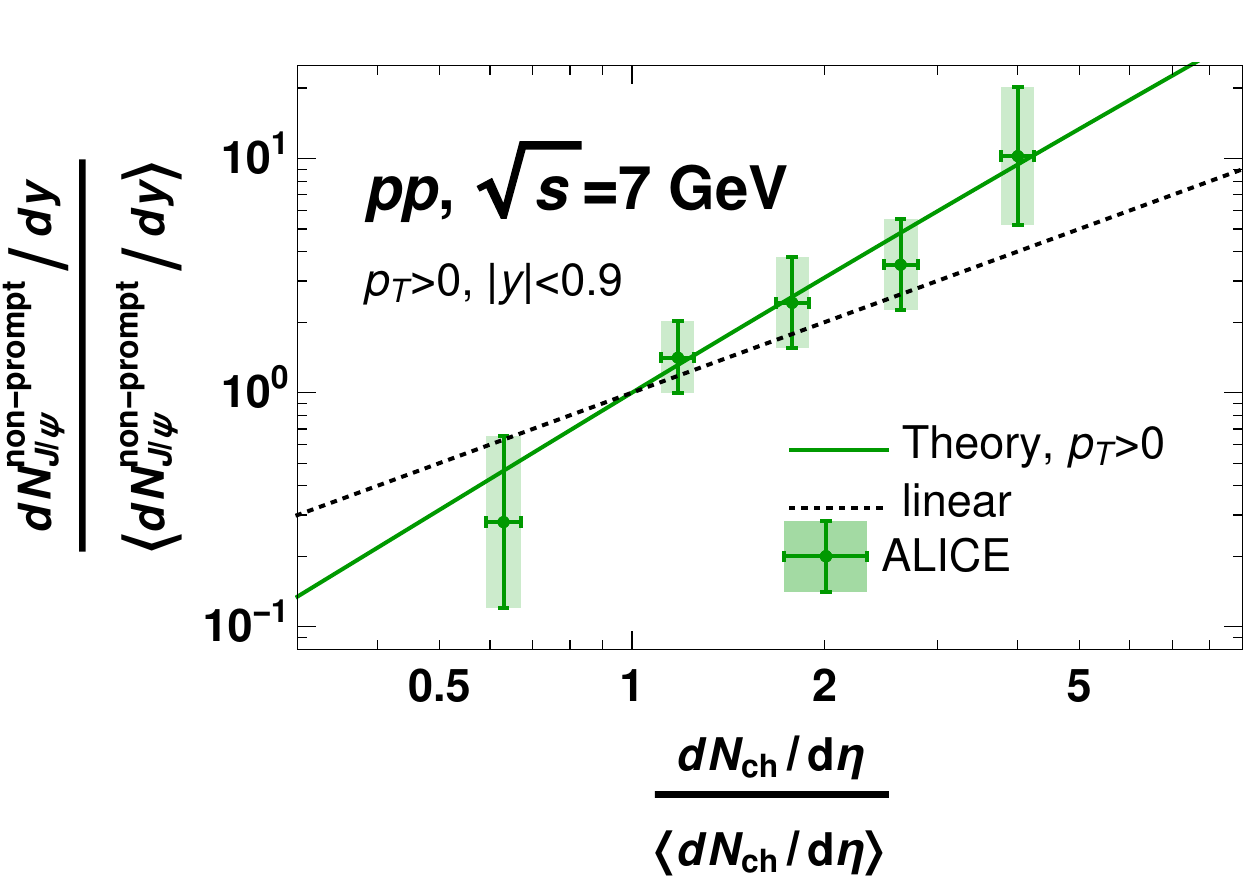}

\caption{\label{DiagsMultiplicityJPsi}Comparison of the theoretical results
for the non-prompt $J/\psi$ mesons with experimental data from ALICE~\cite{Adam:2015ota}.
For the sake of reference we have also shown a short-dashed line which
corresponds to a linear dependence. It is expected that the charged
particles are collected at rapidity window $|y|<0.9$, $|\eta|<1$
.}
 
\end{figure}

Finally, in Figure~\ref{DiagsMultiplicityJPsi} we have shown
the multiplicity dependence of the non-prompt $J/\psi$, which are
formed from the decays of the $b$-mesons. The experimental data clearly
show that the multiplicity dependence grows faster than linear. The
dipole approach provides a very reasonable description of the available
multiplicity dependence.

\section{Conclusions}

\label{sec:Conclusions}In this paper we studied the mechanisms of
the open-heavy flavor meson production. We took into account both
the standard two-pomeron mechanism, as well as estimated the contribution
of three-pomeron fusion. We found that the latter correction is important
for $D$-mesons for small-$p_{T}$ data, where it changes the result
by a factor of two and allows to improve considerably the agreement
of theoretical predictions with data. The correction is less relevant
for $B$-mesons, where it does not exceed ten per cent. As a function
of transverse momentum $p_{T}$, the relative weight of the three-pomeron
contribution decreases, and for $p_{T}\gtrsim10$ GeV the correction
does not exceed one per cent. Such behavior agrees with general expectations
based on large-$p_{T}$ and heavy quark mass limit evaluations. Since
the $p_{T}$-integrated cross-section is dominated by the small-$p_{T}$-region,
we conclude that it is sensitive to the contributions of the three-pomeron
mechanism. Our evaluation is largely parameter-free and relies only
on the choice of the parametrization for the dipole cross-section~(\ref{eq:CGCDipoleParametrization}).

The suggested approach is able to describe the multiplicity dependence
measured by ALICE~\cite{Adam:2015ota}. Contrary to naive expectations,
the relative contribution of the three-pomeron mechanism has a rather complicated
dependence on multiplicity. Due to interplay of direct and interference
contributions, shown in Figure~\ref{fig:NNLOInterference-CutUncut},
the relative contribution of the three-pomeron correction \emph{decreases}
as a function of multiplicity for small $n$, changes sign near $n\approx5$
and starts growing at larger values of $n$. For this reason this
contribution does not lead to a pronounced multiplicity dependence for
the range of multiplicities available from LHC~\cite{Adam:2015ota}.
This result differs dramatically from quarkonia production, where
the measurement of multiplicity dependence was suggested as a means
to estimate the role of the three-pomeron contributions~\cite{Siddikov:2019xvf,Levin:2018qxa}.
This difference occurs due to lack of the interference contributions shown
in the right panel of Figure~\ref{fig:NNLOInterference-CutUncut}, 
in the quarkonia production case.

\section*{Acknowledgements}

We thank our colleagues at UTFSM university and especially Eugene
Levin for encouraging discussions. This research was supported
by ANID PIA/APOYO AFB180002 (Chile) and Fondecyt (Chile) grant 1180232.
Also, we thank Yuri Ivanov for technical support of the USM HPC cluster
where a part of evaluations has been done. 

\appendix

\section{Evaluation of the dipole amplitudes}

\label{sec:Derivation}
\begin{figure}
\includegraphics[width=9cm]{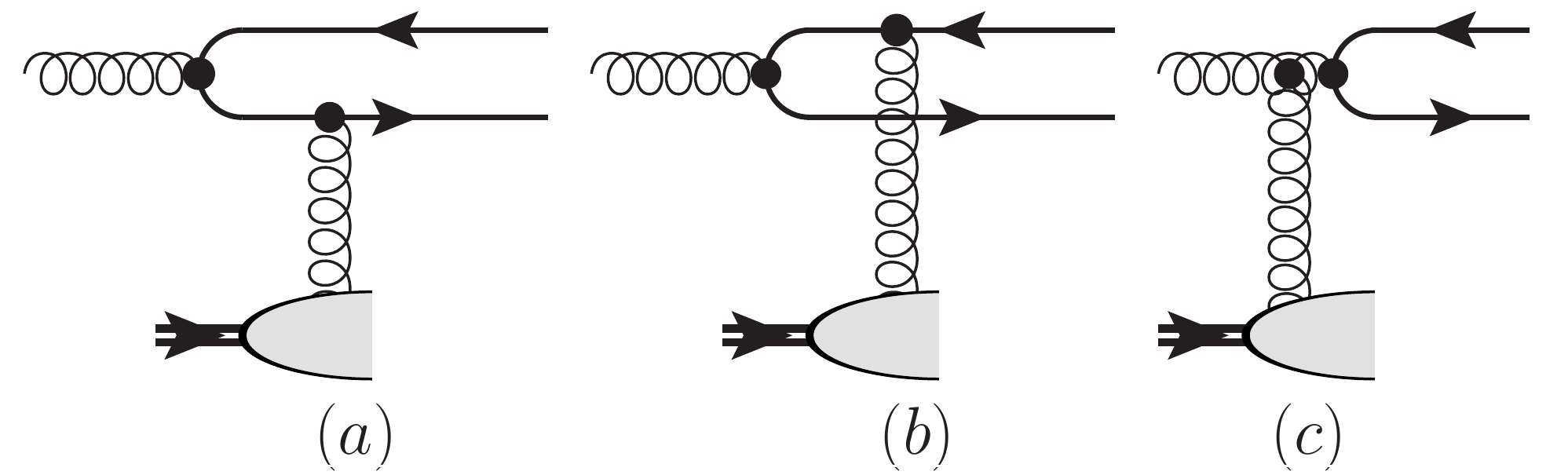}

\caption{\label{fig:Dipole2Pomeron}The diagrams which contribute to the heavy
meson production cross-section in the leading order perturbative QCD.
The contribution of the last diagram ($c$) to the meson formation
can be also viewed as gluon-gluon fusion $gg\to g$ with subsequent
gluon fragmentation $g\to\bar{Q}Q$. In the CGC parametrization of the
dipole cross-section approach, each ``gluon'' is replaced with a reggeized
gluon (BK pomeron), which satisfies the Balitsky-Kovchegov equation
and corresponds to a fan-like shower of soft particles.}
\end{figure}
In this section for the sake of completeness we explain the main technical
steps and assumptions used for derivation of the two-pomeron dipole
cross-section~(\ref{FD1-2},~\ref{eq:N2}) and three-pomeron contributions~(\ref{eq:N3Direct},\ref{eq:N3Interf}).
For heavy quarks it is expected that the strong coupling $\alpha_{s}(m_{Q})$
should be small, which enables the application of perturbative methods.
For this reason it is very instructive to discuss different contributions
in parallel with the perturbative $k_{T}$-factorization-style approach,
tacitly assuming that each gluon should be understood as a parton
shower (``pomeron''). The rules which allow to express the cross-sections
of hard processes in terms of the color singlet dipole cross-section
can be found in~\cite{Kopeliovich:2002yv,Kopeliovich:2001ee}.
In the high-energy eikonal picture, the interaction of the quarks
and antiquark with a target are given by $\pm ig\,t^{a}\gamma\left(\boldsymbol{x}_{\perp}\right)$,
where $\boldsymbol{x}_{\perp}$ is the transverse coordinate of the
quark, and the function $\gamma\left(\boldsymbol{x}_{\perp}\right)$
is related to  a gluonic field of the target. This function is related
to a dipole cross-section $\sigma(x,\,\boldsymbol{r})$ as
\begin{equation}
\Delta\sigma(x,\,\boldsymbol{r})\equiv\sigma(x,\,\infty)-\sigma(x,\,\boldsymbol{r})=\frac{1}{8}\int d^{2}b\left|\gamma\left(x,\,\boldsymbol{b}-z\boldsymbol{r}\right)-\gamma\left(x,\,\boldsymbol{b}+\bar{z}\boldsymbol{r}\right)\right|^{2}\label{eq:DipoleX}
\end{equation}
where $\boldsymbol{r}$ is the transverse size of the dipole, and
$z$ is the light-cone fraction of the dipole momentum carried by
the quarks. The equation~(\ref{eq:DipoleX}) can be rewritten in
the form
\begin{equation}
\frac{1}{8}\int d^{2}\boldsymbol{b}\gamma(x,\,\boldsymbol{b})\gamma(x,\,\boldsymbol{b}+\boldsymbol{r})=\frac{1}{2}\sigma(x,\,\boldsymbol{r})+\underbrace{\int d^{2}b\,\left|\gamma(x,\,\boldsymbol{b})\right|^{2}-\frac{1}{2}\sigma(x,\,\infty)}_{={\rm const}}.\label{eq:SigmaDef}
\end{equation}
For very small dipoles, the dipole cross-section is related to the
gluon uPDF~\footnote{In the literature definitions of the unintegrated PDF $\mathcal{F}\left(x,\,k_{\perp}\right)$
might differ by a factor $k_{\perp}^{2}$.} 
\begin{equation}
\sigma\left(x,\,\vec{\boldsymbol{r}}\right)=\frac{4\pi\alpha_{s}}{3}\int\frac{d^{2}k_{\perp}}{k_{\perp}^{2}}\mathcal{F}\left(x,\,k_{\perp}\right)\left(1-e^{ik\cdot r}\right)+\mathcal{O}\left(\frac{\Lambda_{{\rm QCD}}}{m_{c}}\right),\label{eq:Dip}
\end{equation}
so the functions $\gamma\left(x,\,\boldsymbol{r}\right)$ can be
also related to the unintegrated gluon densities. 

For many high energy processes dominated by the pomeron-pomeron fusion
mechanism it is possible to express the exclusive amplitude or inclusive
cross-section as a sum of the contributions which have the same structure
as the left-hand side of~(\ref{eq:SigmaDef}). For some processes
the last term in~(\ref{eq:SigmaDef}) eventually cancels after summation
over all possible diagrams, so the color dipole density matrix becomes
expressed in terms of the linear combination of the \emph{color singlet}
dipole cross-sections $\sigma(x,\,\boldsymbol{r})$ with different
arguments. While in the deeply saturated regime we can no longer speak
about individual gluons (or pomerons), we expect that the relations
between the dipole amplitudes and color singlet cross-sections should
be valid even in this case. This is a crucial assumption which essentially
constitutes one of the elements of multiplicity dependence discussed
in Section~\ref{subsec:3Pom-1}, and which gives a good description
of the multiplicity dependence~\cite{Siddikov:2019xvf,Levin:2018qxa}. 

For the case of $D$-meson production, the leading-order contribution
is given by the diagrams shown in Figure~(\ref{fig:Dipole2Pomeron})
and yields for the cross-section the result given in~(\ref{FD1-2},~\ref{eq:N2})
(see~\cite{Kopeliovich:2002yv,Kopeliovich:2001ee} for details).
In the evaluation of the $p_{T}$-dependence, we should project the amplitude
in coordinate space (state with definite quark coordinate $r_{Q}$)
onto the state with constant momentum $p_{T}$, by taking an additional
Fourier transform $\int d^{2}\,p_{T}\exp\left(ip_{T}\cdot r\right)$. After
squaring the amplitude in momentum space, this implies
the inclusion into~(\ref{eq:DipoleX}) of the additional factor $\sim\int d^{2}r_{1}d^{2}r_{q}\,e^{ip_{T}\cdot\left(r_{1}-r_{2}\right)}$,
where $\vec{\boldsymbol{r}}_{1,2}$ are the coordinates of the quark
in the amplitude and its conjugate. In the frame where the momentum
of the primordial gluon is not zero, we get an additional convolution
with the $p_{T}$-distribution of the incident (``primordial'')
gluons, as shown in~(\ref{FD1-2}), and was demonstrated in~\cite{Goncalves:2017chx}\footnote{There is a minor difference in the expression for the amplitude~(\ref{eq:N2})
from its analogue which appears in~\cite{Goncalves:2017chx}. The
difference is due to the fact that we request the equality of transverse
coordinates of the heavy antiquarks in the amplitude and its conjugate
instead of the dipole center-of-mass. Besides, in our numerical evaluations
we do not make transition to the momentum space using the linearized
formulas. The reason for this is that for large multiplicity events
such transition might be not justified.}. 

\begin{figure}
\includegraphics[width=9cm]{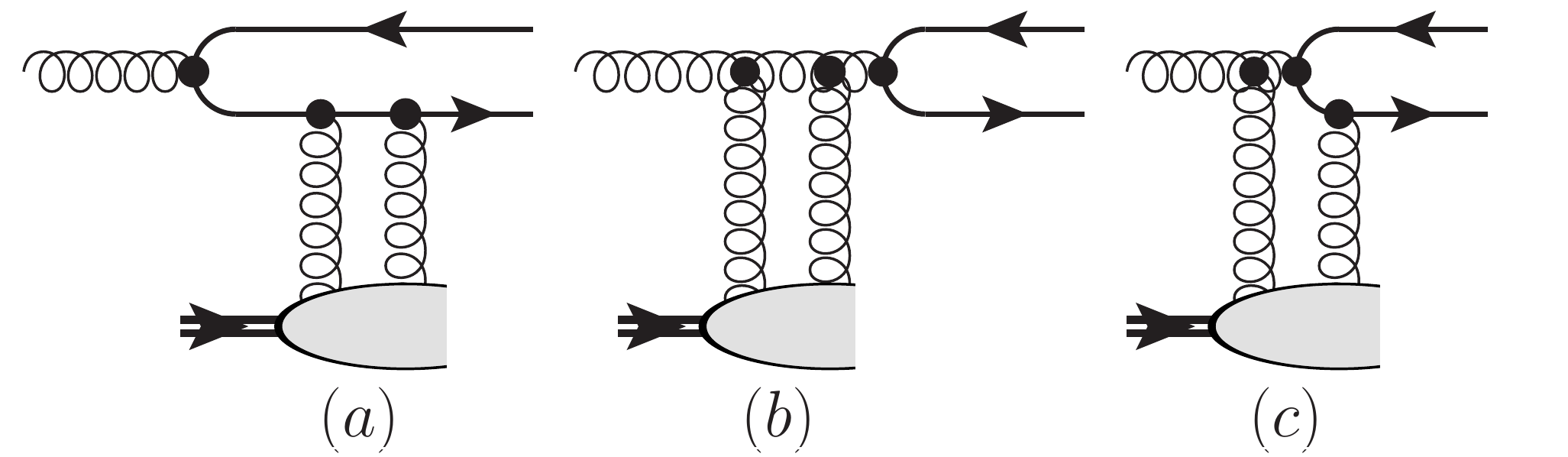}

\caption{\label{fig:Dipole3Pomeron}The diagrams which contribute to the heavy
meson production cross-section in the subleading order in perturbative
QCD ($\mathcal{O}(\alpha_{s})$-correction). In diagrams ($a$) and
$(c)$ all possible attachments of the gluon to the quarks and antiquarks
are implied. In dipole approach each ``gluon'' is replaced with
reggeized gluon (BK pomeron) which satisfies the Balitsky-Kovchegov
equation and corresponds to a fan-like shower of soft particles.}
\end{figure}

For the three-pomeron contribution the above-given approach can
be extended. However, for the description of the interaction with the
target we need to model the multigluon interactions, which are not
taken into account by~(\ref{eq:DipoleX}). In the perturbative limit,
the corresponding interactions are described in terms of the so-called
Double Parton Distribution Functions (DPDFs)~(see~ \cite{Diehl:2011yj,Gaunt:2009re}
for a review and discussion). In general these objects have a complicated
structure, and are not related to the gluon uPDFs. However, at high energies
the correlations between the partons are negligible~\cite{Rinaldi:2013vpa,GolecBiernat:2015aza,Diehl:2013mla},
so the DPDFs can be expressed as products of independent uPDFs.
Thanks to this property, the cross-sections of the so-called Double
Parton Scattering processes can be represented as products of Single
Parton Scattering (SPS) processes. In the Color Glass Condensate
model~\cite{McLerran:1993ni,McLerran:1993ka,McLerran:1994vd} this
assumption is fulfilled automatically. In the color dipole approach, we expect
that the cross-section will be given by a linear combination of structures
$\prod_{k=1}^{4}\gamma(\vec{\boldsymbol{r}}_{k})$, which in view
of~(\ref{eq:DipoleX}) could be separated into a sum of products
of color singlet dipole amplitudes. 

Taking into account all the diagrams shown in Figure~\ref{fig:Dipole3Pomeron},
we obtain for the amplitude of the three-pomeron process

\begin{align*}
\mathcal{A}^{(3)}\left(x,\,\vec{\boldsymbol{r}}_{Q},\,\vec{\boldsymbol{r}}_{\bar{Q}}\right) & =\frac{1}{4}\gamma_{+}\left(x,\,\vec{\boldsymbol{r}}_{Q},\,\vec{\boldsymbol{r}}_{\bar{Q}}\right)\left(d_{ack}\gamma_{-}\left(x,\,\vec{\boldsymbol{r}}_{Q},\,\vec{\boldsymbol{r}}_{\bar{Q}}\right)+if_{ack}\gamma_{+}\left(x,\,\vec{\boldsymbol{r}}_{Q},\,\vec{\boldsymbol{r}}_{\bar{Q}}\right)\right)if_{kbd}t_{d}+\\
 & +\frac{1}{4}\gamma_{-}\left(x,\,\vec{\boldsymbol{r}}_{Q},\,\vec{\boldsymbol{r}}_{\bar{Q}}\right)\left(d_{ack}\gamma_{-}\left(x,\,\vec{\boldsymbol{r}}_{Q},\,\vec{\boldsymbol{r}}_{\bar{Q}}\right)+if_{ack}\gamma_{+}\left(x,\,\vec{\boldsymbol{r}}_{Q},\,\vec{\boldsymbol{r}}_{\bar{Q}}\right)\right)\left(d_{kbd}t_{d}\right)\\
 & +\gamma_{-}^{2}\left(x,\,\vec{\boldsymbol{r}}_{Q},\,\vec{\boldsymbol{r}}_{\bar{Q}}\right)\frac{\delta_{ac}}{6}t_{b}\\
 & +\frac{1}{12}\gamma_{-}\left(x,\,\vec{\boldsymbol{r}}_{Q},\,\vec{\boldsymbol{r}}_{\bar{Q}}\right)\left(d_{acb}\gamma_{-}\left(x,\,\vec{\boldsymbol{r}}_{Q},\,\vec{\boldsymbol{r}}_{\bar{Q}}\right)+if_{acb}\gamma_{+}\left(x,\,\vec{\boldsymbol{r}}_{Q},\,\vec{\boldsymbol{r}}_{\bar{Q}}\right)\right)+\left(b\leftrightarrow c\right)
\end{align*}
where 
\begin{align*}
\gamma_{+}\left(x,\,\vec{\boldsymbol{r}}_{1},\,\vec{\boldsymbol{r}}_{2}\right) & =\gamma\left(x,\,\vec{\boldsymbol{r}}_{1}\right)+\gamma\left(x,\,\vec{\boldsymbol{r}}_{2}\right)-2\gamma\left(x,\,\frac{\vec{\boldsymbol{r}}_{1}+\vec{\boldsymbol{r}}_{2}}{2}\right),\\
\gamma_{-}\left(x,\,\vec{\boldsymbol{r}}_{1},\,\vec{\boldsymbol{r}}_{2}\right) & =\gamma\left(x,\,\vec{\boldsymbol{r}}_{1}\right)-\gamma\left(x,\,\vec{\boldsymbol{r}}_{2}\right),
\end{align*}
$a$ is the color index of the incident (projectile) gluon, $b$ and
$c$ are the color indices of the gluons attached to the target (vertical$t$-channel
gluons in Figure~\ref{fig:Dipole3Pomeron}), $\vec{\boldsymbol{r}}_{Q},\,\boldsymbol{r}_{\bar{Q}}$
are the coordinates of the quarks. For the evaluation of the cross-section
we should square the amplitude, and potentially could get different
structures with $bc\not=b'c'$. Indeed, as was demonstrated in~\cite{Diehl:2011yj,Diehl:2017wew}.
For the Double Parton Distribution Functions (DPDFs) the corresponding
cross-section is described by 6 different color structures, which take
into account possible color state of the gluon pairs in the amplitude
and its conjugate, viz.:
\begin{align}
 & \delta^{bb'}\delta^{cc'},\,f^{bb'k}f^{cc'k},\,d^{bb'k}d^{cc'k},\,t_{10}^{bb',\,cc'},\,t_{27}^{bb',\,cc'},\label{eq:structure}
\end{align}
where 
\begin{align*}
 & t_{10}^{bb',\,cc'}=\delta^{bc}\delta^{b'c'}-\delta^{bc'}\delta^{b'c}-\frac{2}{3}f^{bb'k}f^{cc'k}-i\left(d^{bck}f^{b'c'k}+f^{bck}d^{b'c'k}\right)\\
 & t_{27}^{bb',\,cc'}=\delta^{bc}\delta^{b'c'}+\delta^{bc'}\delta^{b'c}-\frac{1}{4}\delta^{bb'}\delta^{cc'}-\frac{6}{5}d^{bb'k}d^{cc'k}
\end{align*}
However, as was illustrated in~\cite{Korchemsky:2001nx}, the largest
intercept has a configuration when the two gluons are in a relative
color singlet state (two cut pomerons), which corresponds to the first
term in~(\ref{eq:structure}). This configuration dominates at high
energies, and for this reason in what follows we will take into account
only this contribution.

For the evaluation of the $p_{T}$-dependent cross-section  we need to
project the coordinate space quark distribution onto the state with
definite transverse momentum $\boldsymbol{p}_{T}$, and so we have for
the square of the amplitude
\begin{align}
\left|A^{(3)}\left(\boldsymbol{p}_{T}\right)\right|^{2} & =\int d^{2}\boldsymbol{x}_{\bar{Q}}\int d^{2}\boldsymbol{x}_{Q}\int d^{2}\boldsymbol{y}_{Q}\,e^{i\boldsymbol{p}_{T}\cdot\left(\boldsymbol{x}_{Q}-\boldsymbol{y}_{Q}\right)}\,\,\left.\left(A^{(3)}\left(\vec{\boldsymbol{x}}_{i}\right)\right)^{*}A^{(3)}\left(\vec{\boldsymbol{y}}_{i}\right)\right|_{\vec{\boldsymbol{x}}_{\bar{Q}}=\vec{\boldsymbol{y}}_{\bar{Q}}}=\label{eq:Asq}\\
 & =\int d^{2}\boldsymbol{x}_{\bar{Q}}\int d^{2}\boldsymbol{x}_{Q}\int d^{2}\boldsymbol{y}_{Q}\,e^{i\boldsymbol{p}_{T}\cdot\left(\boldsymbol{x}_{Q}-\boldsymbol{y}_{Q}\right)}\times\\
 & =\frac{N_{c}^{2}-1}{4}\left[\gamma_{+}^{2}\left(\vec{\boldsymbol{x}}_{Q},\,\vec{\boldsymbol{x}}_{\bar{Q}}\right)\gamma_{+}^{2}\left(\vec{\boldsymbol{y}}_{Q},\,\vec{\boldsymbol{y}}_{\bar{Q}}\right)\left(\underbrace{\frac{3N_{c}^{2}}{8}}_{27/8}\right)+\right.\nonumber \\
 & +\gamma_{-}^{2}\left(\vec{\boldsymbol{x}}_{Q},\,\vec{\boldsymbol{x}}_{\bar{Q}}\right)\gamma_{-}^{2}\left(\vec{\boldsymbol{y}}_{Q},\,\vec{\boldsymbol{y}}_{\bar{Q}}\right)\left(\underbrace{\frac{\left(43\,N_{c}^{4}-320N_{c}^{2}+720\right)}{72\,N_{c}^{2}}}_{49/24}\right)+\nonumber \\
 & +\left.\frac{\left(N_{c}^{2}-4\right)}{2}\gamma_{+}\left(\vec{\boldsymbol{x}}_{Q},\,\vec{\boldsymbol{x}}_{\bar{Q}}\right)\gamma_{-}\left(\vec{\boldsymbol{x}}_{Q},\,\vec{\boldsymbol{x}}_{\bar{Q}}\right)\gamma_{+}\left(\vec{\boldsymbol{y}}_{Q},\,\vec{\boldsymbol{y}}_{\bar{Q}}\right)\gamma_{-}\left(\vec{\boldsymbol{y}}_{Q},\,\vec{\boldsymbol{y}}_{\bar{Q}}\right)\right]_{\vec{\boldsymbol{x}}_{\bar{Q}}=\vec{\boldsymbol{y}}_{\bar{Q}}}\nonumber 
\end{align}
For the $p_{T}$-integrated cross-section these formulas simplify
since we will have to put $\vec{\boldsymbol{x}}_{i}\equiv\vec{\boldsymbol{y}}_{i},\quad i=Q,\,\bar{Q}.$
As discussed earlier, at high energies the correlations between the
partons are negligible, and the two gluons reggeize independently
and are in color singlet state with respect to each other~\cite{GolecBiernat:2015aza},
so the four-pomeron configuration, up to numerical factor $\sim\sigma_{{\rm eff}}^{-1}\approx\left(20\,{\rm mb}\right)^{-1}$,
can be found from~(\ref{eq:Asq}) applying iteratively the relation~(\ref{eq:DipoleX}).
It is possible to demonstrate that after such procedure we can express
the three-pomeron dipole amplitude in terms of the \emph{color singlet}
dipole cross-sections as given in~(\ref{eq:N3Direct}). The evaluation
of the contribution~(\ref{eq:N3Interf}) follows a similar algorithm,
although we have to take into account that one of the pomerons is uncut, 
and this reason it does not contribute to the growth of the multiplicity.

\section{Fragmentation functions}

\label{sec:FragFunctions} 
\begin{figure}
\includegraphics[width=9cm]{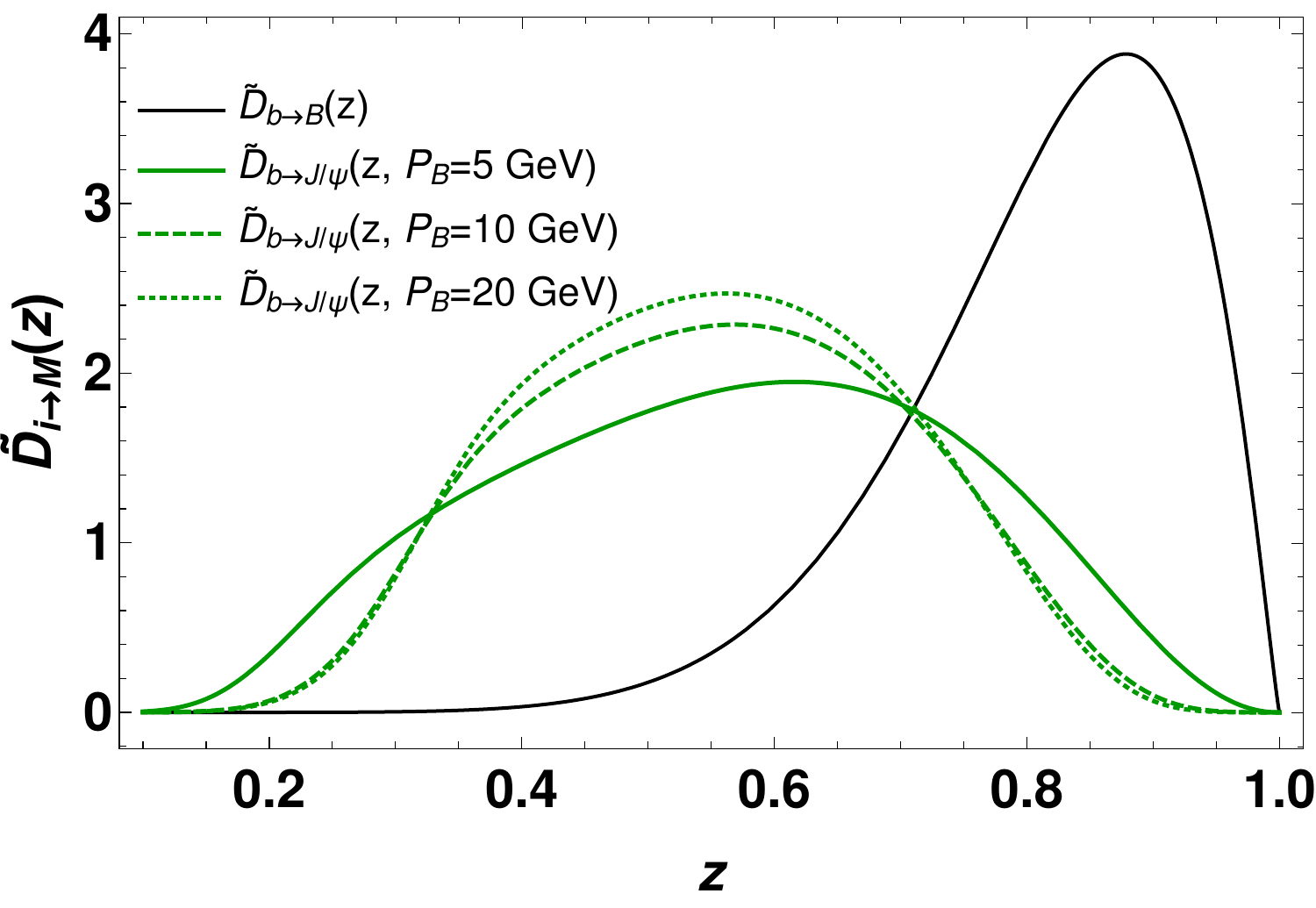}

\caption{\label{fig:fragFunction} The $z$-dependence of the fragmentation
function of $B$-quark cross-section and $J/\psi$ mesons produced
via non-prompt decays of the $B$-mesons, $b\to J/\psi$. For the
ease of comparison we normalized all the fragmentation functions to
unity (so we use the notation $\tilde{D}_{i\to M}$ instead of $D_{i\to M}$).
The normalization coefficients for $b\to B^{\pm}/B^{0}$ and $b\to J/\psi$
cases differ by branching fraction $Br_{B\to J/\psi}\approx0.8\,\%$.
}
\label{DiagsMultiplicity1-1} 
\end{figure}

In this section we would like to briefly
summarize the fragmentation functions used in our evaluations. The
fragmentation functions are nonperturbative objects, which cannot be
evaluated from the first principles. For this reason currently their
parametrization is extracted from the phenomenological fits of $e^{+}e^{-}$
data. For $B$-mesons the dominant contribution comes from the
fragmentation of $b$-quarks, namely one of the four possible subprocesses
$b\to B^{-},\,b\to\bar{B}^{0},\,\bar{b}\to B^{+},\,\bar{b}\to B^{0}$.
If we neglect electroweak corrections, in view of the $u\leftrightarrow d$
flavor symmetry and charge conjugation invariance of QCD, we expect
that the fragmentation functions of all these subprocesses should
coincide. For this reason in what follows we will use just a shorthand
notation for all these processes $b\to B$ . For the fragmentation
function we used the parametrization~\cite{Binnewies:1998vm} 
\begin{equation}
D_{b\to B}\left(z,\,\mu_{0}\right)=N\,z^{\alpha}\left(1-z\right)^{\beta},\label{eq:Db1}
\end{equation}
with the values of free parameters $N=56.4$, $\alpha=8.39$, $\beta=1.16$.
We cross-checked that its predictions are close to results obtained
with the Peterson's parametrization~\cite{Peterson:1982ak}
\begin{align}
D_{b\to B}\left(z,\,\mu_{0}\right) & =\frac{N}{z\left(1-\frac{1}{z}-\frac{\epsilon}{1-z}\right)^{2}},\label{eq:Db2}\\
\epsilon & \approx0.0126
\end{align}
The non-prompt charmonia are produced from decays of the $B$-mesons,
and for this reason their fragmentation function can be related to $D_{b\to B}$
as~\cite{Kniehl:1999vf}
\[
D_{i\to\psi}\left(z,\,\mu\right)=\int_{z}^{1}dx\,D_{i\to B}\left(\frac{x}{z},\,\mu^{2}\right)\times\frac{1}{\Gamma_{B}}\frac{d\Gamma}{dz}\left(z,\,P_{B}\right)
\]
where $\Gamma_{B}\equiv1/\tau_{B}$ is the total decay width of the
$B$-meson, the parameter $P_{B}$ is related to rapidity $y$ and the
transverse momentum $p_{T}$ of the produced charmonium as $P_{B}=\sqrt{p_{T}^{2}+\left(p_{T}^{2}+m_{\psi}^{2}\right)\sinh^{2}y}/z$.
The function $d\Gamma\left(z,\,P_{B}\right)/dz$ was evaluated in
detail in~\cite{Kniehl:1999vf}. Due to space limitations we do not
write out the full expressions for this function and instead in 
Figure~\ref{fig:fragFunction} we compare the fragmentation functions
$D_{b\to B}$ and $D_{b\to J/\psi}$. These two functions differ by
more than two orders of magnitude, and for this reason in order to facilitate
comparison, we plotted the distributions normalized to unity, $\tilde{D}(z)=D(z)/\int_{0}^{1}dz\,D(z)$.
As we can see, the distribution $D_{b\to J/\psi}$ is significantly
wider and has a peak near smaller values of $z\approx0.5$.

For the case of $D$-mesons we should take into account that there
are two complementary mechanisms, a direct (prompt) production, and
indirect (non-prompt) mechanism from decays of $B$-quarks. In both
cases we use a fragmentation function taken from~\cite{Kneesch:2007ey},
\begin{equation}
D_{i\to D}\left(z,\,\mu_{0}\right)=N_{i}\,z^{-\left(1+\gamma_{i}^{2}\right)}\left(1-z\right)^{a}\exp\left(-\gamma_{i}^{2}/z\right),\quad i=b,\,c\label{eq:Dc}
\end{equation}
with parameters given in Table~\ref{tab:Parameters}. Despite
of the significant difference between the values of constants between
$D^{+}$ and $D^{0}$ mesons, the two parametrizations have very similar
shapes and differ only by a factor of two in normalization.

\begin{table}
\begin{tabular}{|c|c|c|c|c|c|c|}
\hline 
 & $N_{c}$ & $a_{c}$ & $\gamma_{c}$ & $N_{b}$ & $a_{b}$ & $\gamma_{b}$\tabularnewline
\hline 
$D^{0}$ & $8.8\times10^{6}$ & $1.54$ & $3.58$ & $78.5$ & $5.76$ & $1.14$\tabularnewline
\hline 
$D^{+}$ & $5.67\times10^{5}$ & $1.16$ & $3.39$ & $185$ & $7.08$ & $1.42$\tabularnewline
\hline 
\end{tabular}\caption{\label{tab:Parameters}The values of parameters used for evaluation
of the $D$-meson fragmentation function with parametrization~(\ref{eq:Dc})
(see~\cite{Kneesch:2007ey} for details).}
\end{table}

 \end{document}